\documentclass[manuscript]{aastex6}
\usepackage{graphicx}   % needed for figures
\usepackage{dcolumn}    % needed for some tables
\usepackage{bm}         % for math
\usepackage{verbatim}   % for math
\usepackage{booktabs}
\usepackage{amsmath,amssymb}
\usepackage{color}
\usepackage{multirow}
\usepackage{inputenc}
\usepackage{natbib}
\bibpunct{(}{)}{;}{a}{}{,}
\usepackage{floatrow}
\usepackage[caption=false]{subfig}
\usepackage{savesym}
\savesymbol{tablenum}
\usepackage{siunitx}
\restoresymbol{SIX}{tablenum}
\usepackage{enumitem}   % custom enumerate labels

\usepackage[normalem]{ulem}
\newcommand\bsout{\bgroup\markoverwith{\textcolor{green}{\rule[0.5ex]{2pt}{0.4pt}}}\ULon}

\newcommand{\aver}[1]{\ensuremath{\left\langle#1\right\rangle}}
\renewcommand{\d}{\mathrm d}

\begin{document}
\title{Stochastic Fermi Energization of Coronal Plasma during explosive magnetic energy release} 
\author{Theophilos Pisokas, Loukas Vlahos, Heinz Isliker and Vassilios Tsiolis}
\affiliation{Department of Physics, Aristotle University of Thessaloniki\\
GR-52124 Thessaloniki, Greece}
\author{Anastasios Anastasiadis}
\affiliation{Institute for Astronomy, Astrophysics, Space Applications and Remote Sensing,\\
National Observatory of Athens\\
GR-15236 Penteli, Greece } 
\date{\today}
\begin{abstract} 
The aim of this study is to analyze the interaction of charged particles (ions and electrons) with randomly formed particle scatterers (e.g.\ large scale local ``magnetic fluctuations'' or ``coherent magnetic irregularities'' ), using the set up proposed initially by \cite{Fermi49}. These scatterers are formed by the explosive magnetic energy release and propagate with the
Alfv\'en speed along the irregular magnetic fields.  They are large scale local fluctuations  ($\delta B/B \approx 1$), randomly distributed inside the unstable magnetic topology and will here be called 
%in this article 
 {\bf Alfv\'enic Scatterers (AS)}. We constructed a 3D grid on which a small fraction of randomly chosen grid points are acting as AS. In particular, we study how a large number of test particles evolve inside a  collection of AS,
%. We 
analyzing 
%the energy gain of the particles, 
the evolution of their energy distribution and their escape time distribution. We use a well established method 
%\citep{Ragwitz2001} 
to estimate the transport coefficients directly from the trajectories of the particles.
% interacting with the scatterers. 
Using the estimated transport coefficients and solving the Fokker-Planck (FP) equation numerically, we can recover the energy distribution of the particles. We have shown that the Stochastic Fermi Energization (SFE) of mildly relativistic and relativistic plasma can heat and accelerate the tail of  the ambient particle distribution as predicted by \citet{Parker58} and \citet{Ramaty79}. The temperature of the hot plasma and the tail of the energetic particles depend on the mean free path ($\lambda_{\rm sc}$) of the particles between the scatterers inside the energization volume. 
\end{abstract}
\maketitle
%-----------------------------------------------------------------------------------
%-----------------------------------------------------------------------------------
\section{Introduction}
The stochastic acceleration was first proposed and analyzed by \cite{Fermi49} as a mechanism for the acceleration of Cosmic Rays \cite[see details and references therein]{Longair11}. The core of his idea, which we call here Stochastic Fermi Energization (SFE),  had a larger impact on non-linear processes in general and has been the driving force behind all subsequent theories on charged particle energization in space and astrophysical plasmas. In the original treatment, relativistic particles were accelerated by encounters with very massive, slowly moving magnetic clouds (scattering centers). Fermi also dropped (without any justification) the stochastic energy change and kept only the systematic rate of energy gain, focusing on the interaction of scatterers with the high energy tail of protons. The rate of the systematic energy gain of the charged particles with the scatterers is proportional to the square of the ratio of the magnetic cloud speed ($V$) to the speed of light ($c$), i.e.~$(V/c)^2$. A few years after the initial article by Fermi, \cite{Davis} and \cite{Parker58} emphasized the stochastic nature of the initial Fermi proposal and they estimated analytically the transport coefficients, using an idealised assumption for the interaction of the scatterers with the particles. \cite{Parker58, Ramaty79} assumed that the scattering centers are randomly moving hard spheres and applied their model to solar flares, for accelerating protons from the thermal distribution.

The initial idea put forward by Fermi was soon replaced in the astrophysical literature with a new suggestion based on the interaction of charged particles with a Kolmogorov spectrum of {\bf low amplitude} MHD waves $(\delta B/B\ll 1)$  and the acceleration process was renamed as {\bf stochastic (weak) turbulent heating and acceleration} or simply {\bf stochastic acceleration by turbulence (SAT)} \citep{Davis, Tverskoi67, Kulsrud71};  \cite[see also the reviews by][]{Miller97, Petrosian12}. When the amplitude of the waves $(\delta B)$ is much smaller than the mean magnetic field $B$, the transport coefficients for the SAT are estimated with the use of the quasilinear approximation, and by solving the transport equations one can estimate the evolution of the energy distribution of the particles \citep[see][]{Achterberg81,Schlickeiser89}. So, the SFE was replaced in the recent years by SAT, and the Fokker-Planck equation became the main tool for the analysis of the evolution of energy distributions of particles.

In the solar atmosphere, the formation of scatterers from traveling large scale Alfv\'en wave packets, deformed by irregular magnetic fields inside a complex magnetic topology and driven by the turbulent convection zone and/or the photospheric motions, has been analysed by \cite{Sudan89}. They have shown that complicated magnetic geometry, typical to most astrophysical plasmas, greatly reduces the dissipation length, as compared to laminar fields. \citet{Parker94} pointed out that the spontaneous formation of magnetic discontinuities inside a driven complex magnetic topology can cause explosive events (nanoflares, microflares, flares  and Coronal Mass Ejections (CME)), which can become the source of torsional Alfv\'en waves propagating along the mean magnetic field, developing increasingly complex structures when they propagate along the stochastic magnetic field lines.  Hence, the onset of flares of all scales arises in the strongly interwoven flux bundles and drives Alfv\'en wave packets \citep{Parker83}. \cite{Fletcher08} suggested that the energy stored in preflare coronal magnetic field can be rapidly converted and liberated by coronal magnetic field reconfiguration and relaxation during flares. The sudden reconfiguration will generate large scale wave pulses, which transport energy rapidly through the corona and the lower atmosphere. Their analysis was restricted to laminar magnetic fields and simple magnetic loops, and excluded the formation of AS  and unstable current sheets (UCS), which are expected if the large scale MHD waves propagate along complex field lines.

The majority of the 3D MHD simulations of explosive solar energy release start with a simple magnetic topology which is forced away from equilibrium. During the eruption or shortly after, the magnetic field becomes complex and liberates large scale magnetic disturbances due to its rapid reconfiguration. The literature on this topic is vast, since the avenues for the explosive events are several, e.g.~stochastic shuffling (braiding) of the magnetic fields at the foot-points of loops \citep{Galsgaard96}, magnetic flux emergence \citep{Archontis08,Schmieder14}, {\bf  loss of stability of coronal loops \citep{Gerrard03, Gordovskyy2011, Gordovskyy2012, Amari13,Leake14, Torok14, Fletcher15}}. The important points to stress here are: (1) The explosive solar events are associated with complex magnetic topologies; (2) The reconfiguration of the large scale magnetic topology, due to reconnection,  will drive large scale MHD disturbances propagating inside the 3D unstable structure; (3) The entire unstable magnetic topology will form AS and UCS  which participate in the heating and acceleration of the solar plasma trapped by this structure, as we will show in this article.

In this article, we pose three fundamental questions: (1)~Is the SAT a good approximation for the SFE~? (2)~Is the simple expression for the escape time, $t_{\rm esc}=L/v$ (where $L$ is the characteristic size of the acceleration volume and $v$ the characteristic velocity of the accelerated particles), used extensively in the current literature, a valid approximation for the SFE~? (3)~How do trapped particles (in closed or open simulation boxes) evolve during the SFE~?

In this study, we assume that coherent magnetic structures traveling with the Alfv\'en speed  ({\bf Alfv\'enic Scatterers}) can be formed by the interaction of the large amplitude magnetic fluctuations propagating along the complex magnetic topology inside the solar atmosphere. Their excitation can be either through the motion of the plasma inside the convection zone and/or the photosphere, by emerging flux, or by sudden energy release during large scale magnetic reconnection in the solar corona. We concentrate in this article on the interaction of the plasma with AS. The interaction of  electrons and ions with UCS was discussed briefly by \cite{Vlahos16}.

The outline of the present work is as follows: In the next section we recapitulate the main assumptions of the stochastic Fermi type mechanism for particle acceleration and heating, and compare the main results with the stochastic weakly turbulent acceleration. In Section~\ref{s:model} we construct a model based on a 3D lattice approach, where scatterers can easily be replaced by different types of local accelerators, and we apply this model to the solar corona where the ``scatterers'' are local
%mean
fluctuations, called here AS. The simulation box is assumed to be open or with periodic boundary conditions, and the role of collisions is explored. In Section~\ref{s:discussion} we discuss the importance of our results in the context of the heating and acceleration of particles in the solar corona, and in Section~\ref{s:summary} we summarize the main results of our study.

%-----------------------------------------------------------------------------------
%-----------------------------------------------------------------------------------
\section{The critical assumptions of the SFE and its relation with SAT}\label{s:assumptions}
\cite{Fermi49} based the proposed acceleration mechanism
and his estimates on several assumptions. Let us briefly discuss these assumptions, since some of them are not obvious in the current literature \citep[see][]{Longair11}:
\begin{enumerate}[label=(\textbf{\arabic*})~]
    \item The particles move with relativistic velocity $u$ and the scatterers (``magnetic clouds'') move with mean speed $V$, much smaller than the speed of light;
    \item The energy gain or loss of the particles interacting with the scatterers is
        \begin{equation}\label{energyF}
	        \frac{\Delta W}{W}\approx\frac{2}{c^2}(V^2-\vec{V} \cdot \vec{v}) ,
         \end{equation}
    where for head on collisions $\vec{V} \cdot \vec{v}<0$ and the particles gain energy, for overtaking collisions $\vec{V} \cdot \vec{v}>0$ and the particles lose energy ($\vec{v} = \vec{u}/\gamma$, with $\gamma$ the Lorenz factor);
    \item The rate of energy gain is estimated from the relation
        \begin{equation}\label{rate}
	        \frac{\d W}{\d t}=\frac{W}{t_{\rm acc}}=aW,
        \end{equation}
    where
        \begin{equation}\label{timeacc}
	        a=\frac{1}{t_{\rm acc}}=\frac{4}{3} \left(\frac{V}{c} \right)^2 \left( \frac{c}{\lambda_{\rm sc}} \right),
	    \end{equation}
    and $\lambda_{\rm sc}$ is the mean free path the particles travel between the scatterers;
    \item Assuming that the distribution of the scatterers is uniform inside the acceleration volume with density $n_{\rm sc}$, the mean free path will be $\lambda_{\rm sc} \approx \frac{1}{ \sqrt[3]{n_{\rm sc}}}.$ The assumption that the scatters are uniformly distributed in space is a {\bf strong} assumption in turbulent systems. Turbulent systems tend to be highly anisotropic and so also the distribution of the scatterers may not be isotropic;
    \item The particles are not trapped inside the scatterers, their interaction is instantaneous;
    \item Solving eq.~\eqref{rate}, we can estimate the temporal evolution of the mean energy as
        \begin{equation}\label{Energy}
	        \aver{W(t)} = W_0 e^{at};
        \end{equation}
    \item \cite{Fermi49} used the diffusion (Fokker-Planck) equation in order to estimate the evolution of the energy distribution $n(W,t)$ of the accelerated particles. In order to simplify the diffusion equation, he assumed that spatial diffusion is not important and the particles diffuse only in energy space,
        \begin{equation} \label{diff}
            \frac{\partial n}{\partial t} +
            \frac{\partial}{\partial W} \left[F n -\frac{\partial (D n) }{\partial W} \right] =
            -\frac{n}{t_{\rm esc}} + Q ,
        \end{equation}
    where $t_{\rm esc}$ is the escape time from an acceleration volume with characteristic length $L$, $Q$ is the injection rate, $F$ and  $D$ are the transport coefficients.
\end{enumerate}

Fermi reached his famous result by assuming that: (a)~the energy distribution has reached an asymptotic state before the particles escape from the acceleration volume, and (b)~the energy diffusion coefficient approaches zero $(D \sim 0)$ asymptotically for {\bf relativistic particles} and the acceleration is mainly due to the systematic acceleration, which, according to eq.~\eqref{rate}, is given as
\begin{equation}\label{convection}
	F(W,t) = \aver{\frac{\d W}{\d t}}_W = aW.
\end{equation}
Based on all the assumptions mentioned above, the asymptotic solution of eq.~\eqref{diff} is simply
\begin{equation}\label{diff1}
	n(W)\sim W^{-k},
\end{equation}
where
\begin{equation}
    k = 1 + \frac{t_{\rm acc}}{t_{\rm esc}}.
\label{index}
\end{equation}
The index $k$ approaches 2, which is close to the observed value for the cosmic ray spectrum, only if $t_{\rm acc} \approx t_{\rm esc},\; \; a t_{esc} \sim 1$. In most recent theoretical studies of the stochastic Fermi acceleration, the escape time (which is so crucial for the estimate of $k$) is difficult to estimate quantitatively \citep{Miller90, Petrosian12}.

\citet{Parker58} and \citet{Ramaty79} analyzed the interaction of electrons and ions with large amplitude magnetic perturbations, which they assumed to be {\bf hard spheres} in order to obtain analytical results. They estimated the transport coefficients analytically as 
\begin{equation} \label{conParker}
F(W)= \frac{4}{3} \alpha c \left [ \frac{2W}{m} \right]^{1/2} \sim W^{0.5}
\end{equation}
and 
\begin{equation} \label{diffParker}
	D(W)=\frac{1}{3}\alpha c\left [ 2W\right ]^{3/2} m^{1/2}\sim W^{1.5},
\end{equation}
with $\alpha = (3/4) a.$
The mean energy increase in the hard sphere approximation  is 
\begin{equation} \label{meanhs}
	 <W(t)> \sim t^2.
\end{equation}
The energy distribution is obtained as an analytical solution of the Fokker Planck equation, and for low energy particles $(W<<mc^2)$, it can be approximated with the function

\begin{equation} \label{leparticles}
	f(W) \sim K_2\left ( 2 \sqrt{\frac{3p}{mc \; \alpha t_{esc}}}  \right ) ,
\end{equation} 
where $K_2$ is the Bessel function of the second kind and $p$ the momentum of the particles. For relativistic particles ($W>>mc^2$), the solution is 

\begin{equation} \label{heparticles}
	f(W) \sim W^{1/2- (1/2)(9+12/(\alpha\; t_{esc}))^{1/2}} .
\end{equation} 
Therefore the results reported by Fermi in his original article should be modified for non relativistic  or relativistic particles, when the analysis is based on the assumption of hard spheres. 
Eqs. (\ref{heparticles}),  (\ref{diff1}) and (\ref{index}) yield the same result for $a \; t_{esc} \sim 1.$ 

Let us now summarize the main results from the interaction of a spectrum of isotropic Alfv\'en waves with the ambient ions. If we assume that the spectral density $W_k$ (energy density per unit wavenumber) is proportional to $k^{-2},$ we can estimate the acceleration time \citep[see][]{Miller90} as
\begin{equation}
	\frac{1}{t_{\rm acc}} \approx \left( \frac{3\pi}{64} \right) \left( \frac{V_A}{c} \right)^2 \left( \frac{U_A}{U_B} \right) (c\;k_{\min}) ,
\end{equation}
where $V_A$ is the Alfv\'en speed, $k_{\min}$ is the minimum wave number, $U_A$ is the total energy density in the Alfv\'en waves (obtained by integrating $W_k$ from $k_{\rm min}$ to infinity), and $U_B$ the energy density of the ambient magnetic field. For typical coronal parameters the ambient magnetic field is \SI{100}{G}, for weak turbulence $(U_A/U_B) \approx 10^{-2}$, and with $k_{\min} \approx \SI{3e-6}{cm^{-1}}$ for the Alfv\'en waves to be in resonance with ions with energy as high as \SI{10}{GeV}, we estimate $t_{\rm acc} \approx \SI{15}{sec}$ and, assuming that the waves act as Fermi scatterers (see eq.~\ref{timeacc}), their $\lambda_{\rm sc} \approx \SI{2e8}{cm}$. Therefore,  stochastic weak Alfv\'enic turbulence, if everything is fine-tuned, can provide $t_{\rm acc}$ and $\lambda_{\rm sc}$  of the ions during solar flares, but it fails to provide information on the energy distribution, since the escape time is qualitatively defined through a simple ballistic relation $t_{\rm esc} \approx L/v$, where $L$ is the length of the acceleration volume.

\cite{Hamilton92} analysed the resonant interaction of a very weak spectrum of whistler waves with electrons in the presence of collisions for trapped ($t_{\rm esc} \longrightarrow \infty$) and non trapped electrons by using and solving the Fokker Planck equation. They found that, for energies $E<E_c$ the particles are influenced by collisions and form a quasithermal distribution, whereas above $E_c$ the distributions of the energetic particles are power laws. The energy $E_c$ depends on the conditions of the ambient plasma.  They have also shown that, even for trapped particles ($t_{\rm esc} \longrightarrow \infty$), the interaction of the electrons with the waves reaches an asymptotic state, which is similar to the one obtained from the ``leaky'' acceleration volumes.

\cite{Greco10} attempted a numerical study of Fermi acceleration by using a 2D model, where ions can experience a Fermi-like acceleration processes by interacting with a synthetic oscillating electromagnetic wave, which is carefully tailored to mimic magnetic fluctuations (magnetic clouds), randomly positioned within the $xy$-plane. The free parameter of their system is the mean free path between the ``magnetic clouds'', $\lambda_{\rm sc}$. Their results show an efficient heating of the ambient ions, especially when the magnetic clouds are densely packed and $\lambda_{\rm sc}$ is small. In their analysis no attempt was made to estimate the transport coefficients, or to provide any comparison with the basic characteristics of the  stochastic Fermi acceleration for the high energy part of their distribution (e.g. power-law index of the accelerated particles, acceleration time, evolution of the mean energy etc).

A key element in our understanding of stochastic Fermi acceleration is the careful estimation of the transport coefficients and the detailed analysis of the escape time {\bf for non-trapped particles}. We have therefore made an attempt to calculate both in detail in this article. The energy diffusion coefficient can be estimated directly from the dynamics of the particles through the relation
\begin{equation}\label{eq:DWW}
	D(W,t) = \frac{\aver{\left(W(t+\Delta t) - W(t)\right)^2}_W}{2\Delta t},
\end{equation}
and the energy convection coefficient, representing the systematic acceleration, is given as
\begin{equation}\label{eq:FW}
	F(W,t) =\frac{\aver{W(t+\Delta t) - W(t)}_W}{\Delta t}.
\end{equation}
Here, $\aver{\ldots}_W$ denotes the conditional average that $W(t) = W$, which is applied in order to determine the functional dependence of the transport coefficients on
the energy $W$ \cite[see e.g.~][]{Ragwitz2001}. In practice, the energies
of the particles at time $t$
are divided into bins of finite size, and the transport coefficients are determined for each of the subsets of particles in the bins. $\Delta t$ must be a small time-interval, which should just be large enough so that most particles show measurable changes of the energy over the time interval
$\Delta t$ (theoretically the limit $\Delta t \to 0$ would apply).

%-----------------------------------------------------------------------------------
%-----------------------------------------------------------------------------------
\section{Alfv\'enic Scatterers}\label{s:model}
\subsection{Initial set-up}
We construct a 3D grid $(N \times N \times N)$ with linear size $L$, with grid size $\ell=L/(N-1)$. Each grid point is set as either \emph{active} or \emph{inactive}, i.e.~a scatterer or not. Only a small fraction $R = N_{\rm sc}/N^3$ of the grid points are active (5-\SI{15}{\percent}). We can define the density of the scatterers as $n_{\rm sc} = R \times N^3/L^3$,
and the mean free path of the particles between scatterers can be determined as $\lambda_{\rm sc}=\ell/R.$
When a particle (an electron or an ion) encounters an active grid point, it renews its energy state depending on the physical characteristic of the scatterer. It then moves in a random direction with its renewed velocity $v$, until it meets another active point or exits the grid. The minimum distance between two scatterers is the grid size ($\ell$). The time between two consecutive scatterings is $\Delta t = s/v,$ where $s$ is the distance the particle travels, which is an integer multiple of the minimum distance $\ell$.

At time $t = 0$ all particles are located at random positions on the grid. The injected distribution $n(W, t=0)$ is a Maxwellian with temperature $T$. The initial direction of motion of every particle is selected randomly.

%-----------------------------------------------------------------------------------
%\subsection{Random ``scattering'' by magnetic clouds}
\subsection{Open boundary conditions}\label{Open}
Here we are using the standard stochastic Fermi energization process, as discussed in the previous section. The parameters used in this article are related to the plasma parameters in the low solar corona. We choose the strength of the magnetic field to be $B = \SI{100}{G}$, the density of the plasma $n_0 = \SI{e9}{cm^{-3}}$, the ambient temperature around \SI{100}{eV}, and the length $L$ of the simulation box is \SI{e10}{cm}. The Alfv\'en speed is $V_A \approx \SI{7e8}{cm/sec}$, so $V_A$ is comparable with the thermal speed of the electrons. Each scattering changes the energy of a particle according to eq.~\ref{energyF}. As the lattice constrains the motion, the term $\vec{V} \cdot \vec{v}$ can randomly assume only three possible values: -1 for head on scatterings, 1 for overtaking scatterings, and 0 for perpendicular scatterings. The typical energy increment is of the order of $(\Delta W/W) \approx (V_A/c)^2\sim \num{e-4}$. We consider the grid to be open, so particles can escape from the acceleration region when they reach any boundary of the grid, at $t = t_{\rm esc}$, which of course is different for each escaping particle. We assume in this set-up that only $R = \SI{10}{\percent}$ of the $N^3 = 601^3$ grid points are active.

\begin{figure}[htb]
	\sidesubfloat[]{\includegraphics[width=0.40\columnwidth]{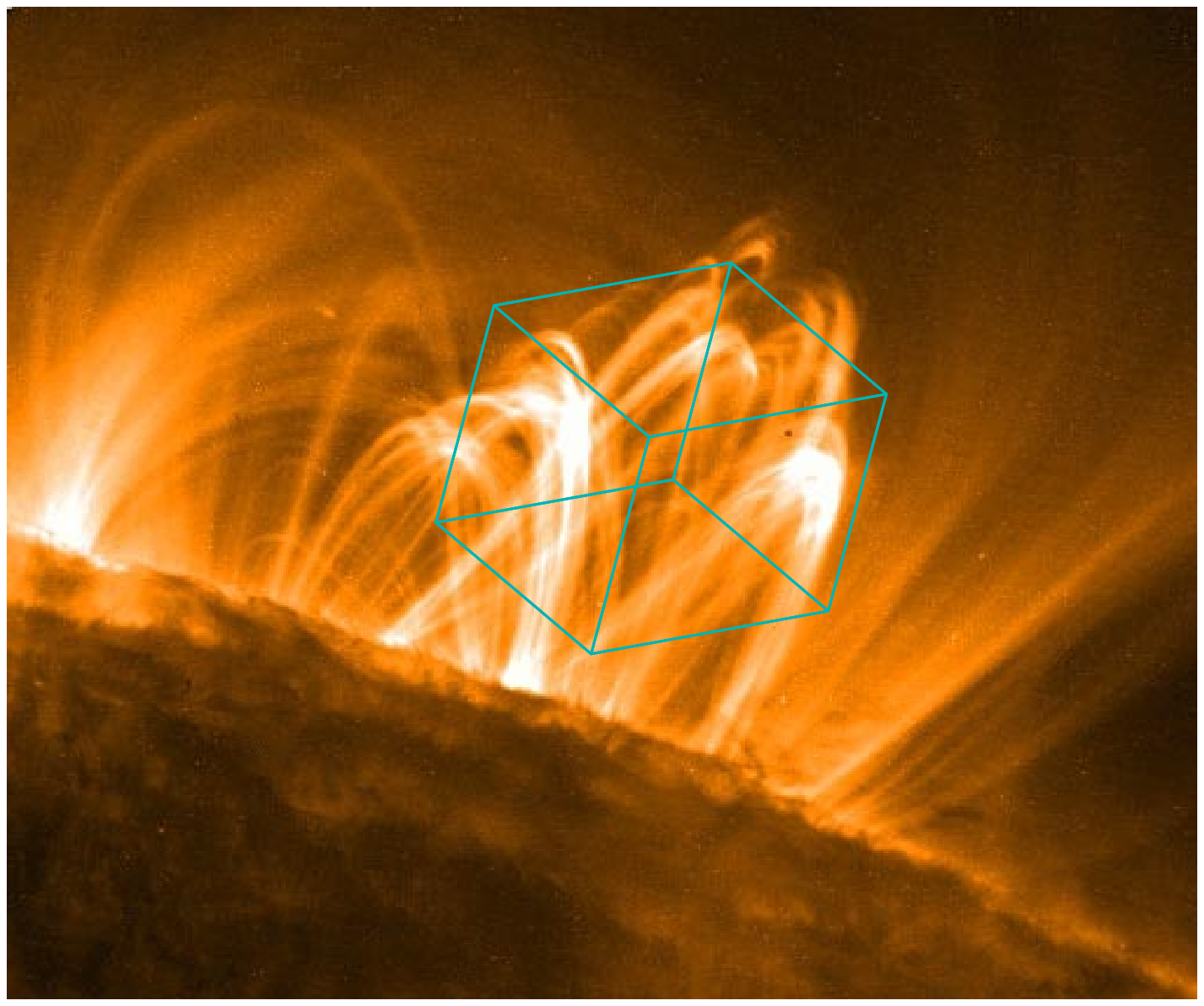}%
		\label{f:3dBox:loop}}\hfill%
	\sidesubfloat[]{\includegraphics[width=0.38\columnwidth]{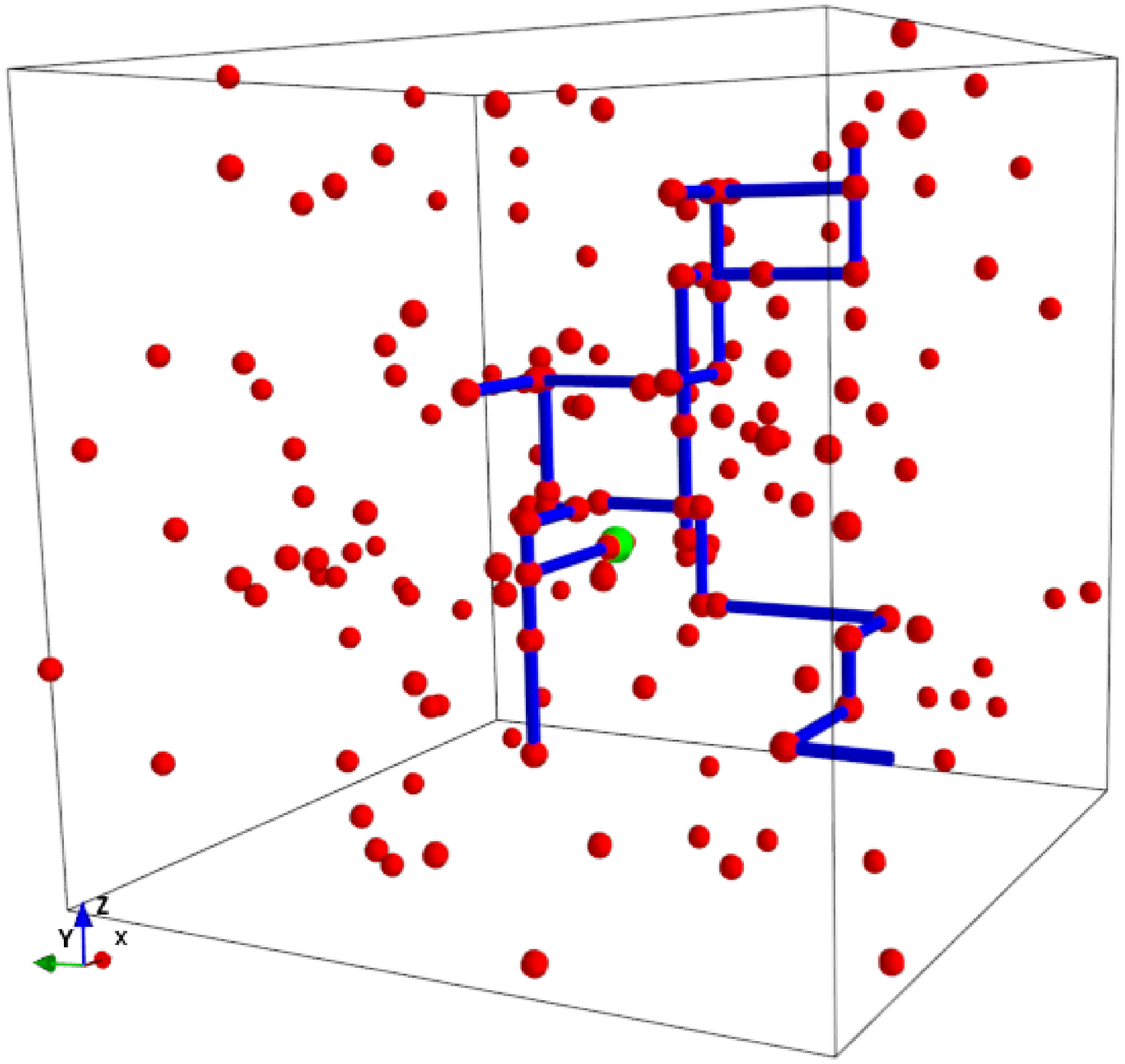}%
		\label{f:3dBox:box}}%
	\caption{\protect\subref{f:3dBox:loop} The complex magnetic topology inside an active region, if it is disturbed by a magnetic perturbation, forms Alfv\'enic Scatters, which are  one of the  basic processes for  coronal heating and particle acceleration during explosive  energy release. \protect\subref{f:3dBox:box} The trajectory of a typical particle (blue tube) inside a grid with linear dimension $L=\SI{e10}{cm}$. Active points are marked by spheres in red color. The particle starts at a random grid-point (green sphere), moves along a straight path on the grid till it meets an active point and then it moves into a new random direction, and so on, until it exits the simulation box.}\label{f:3dBox}
\end{figure}

The AS are formed through the propagation of large magnetic disturbances inside a complex magnetic topology (see Fig.~\ref{f:3dBox:loop}). A typical trajectory of a particle inside the simulation box is displayed in Fig.~\ref{f:3dBox:box}, the particles move along the grid on straight lines until they encounter a scatterer, which affects their energy and direction of motion (see eq.~\ref{energyF}). The motion of the particles is typical for a stochastic system with random-walk like gain and loss of energy before exiting the simulation box. The mean free path is calculated as $\lambda_{\rm sc} = \ell/R \approx \SI{1.7e8}{cm}$, which coincides with the value estimated numerically by tracing particles inside the simulation box.

\begin{figure}[ht]
	\sidesubfloat[]{\includegraphics[width=0.40\columnwidth]{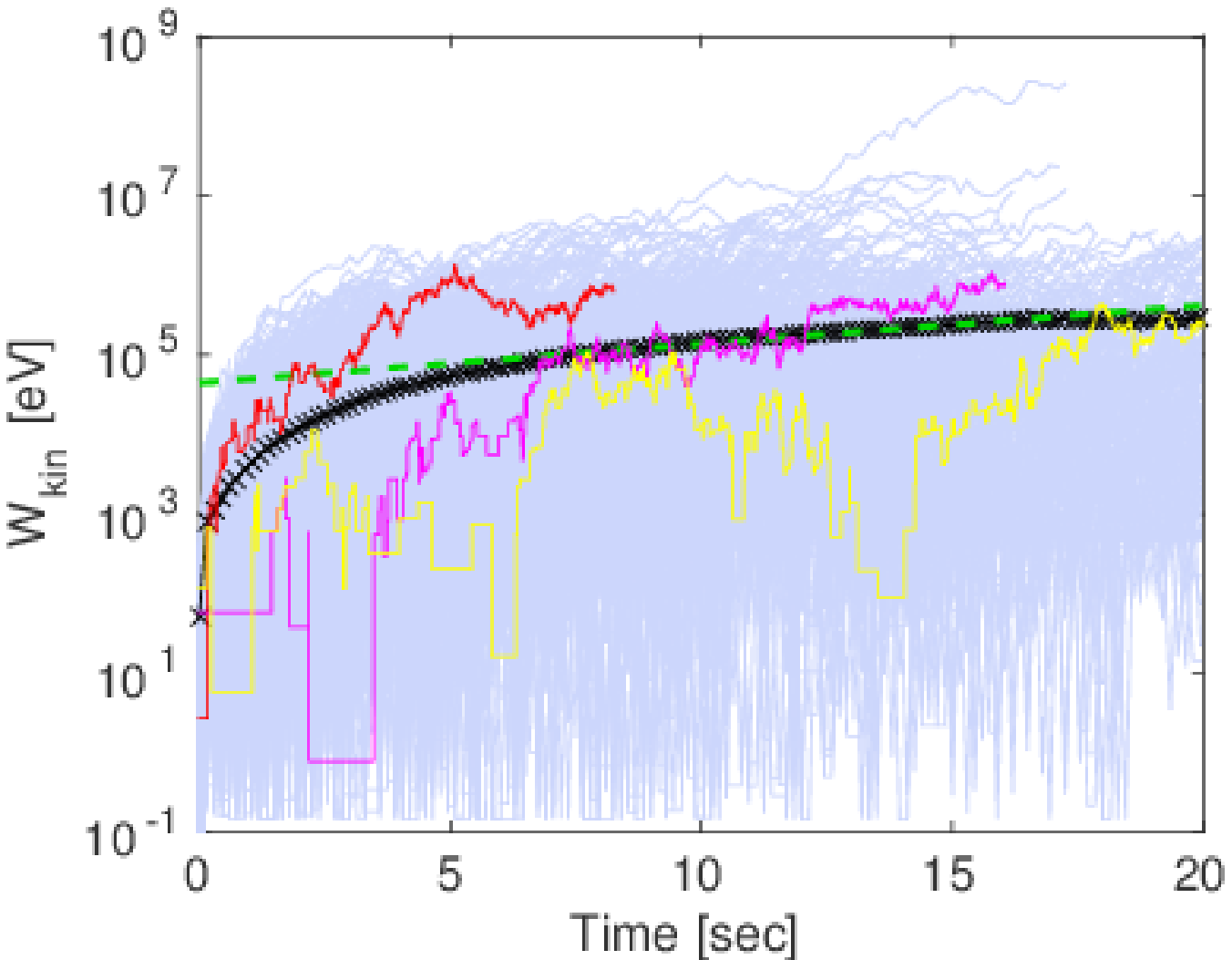}%
        \label{f:F2o:mW}}\hfill
    \sidesubfloat[]{\includegraphics[width=0.40\columnwidth]{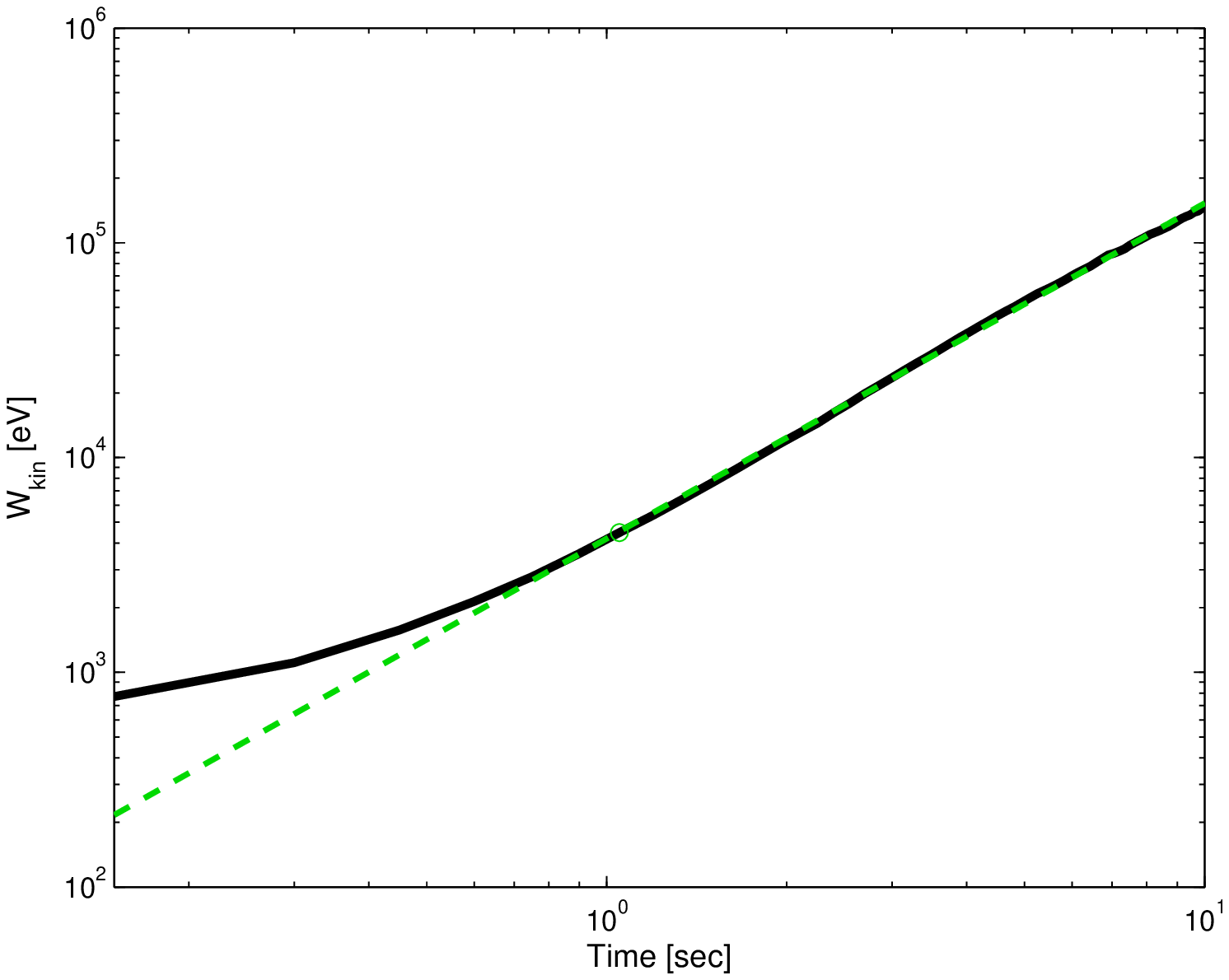}%
        \label{f:F2o:mWlog}}\\
	\sidesubfloat[]{\includegraphics[width=0.40\columnwidth]{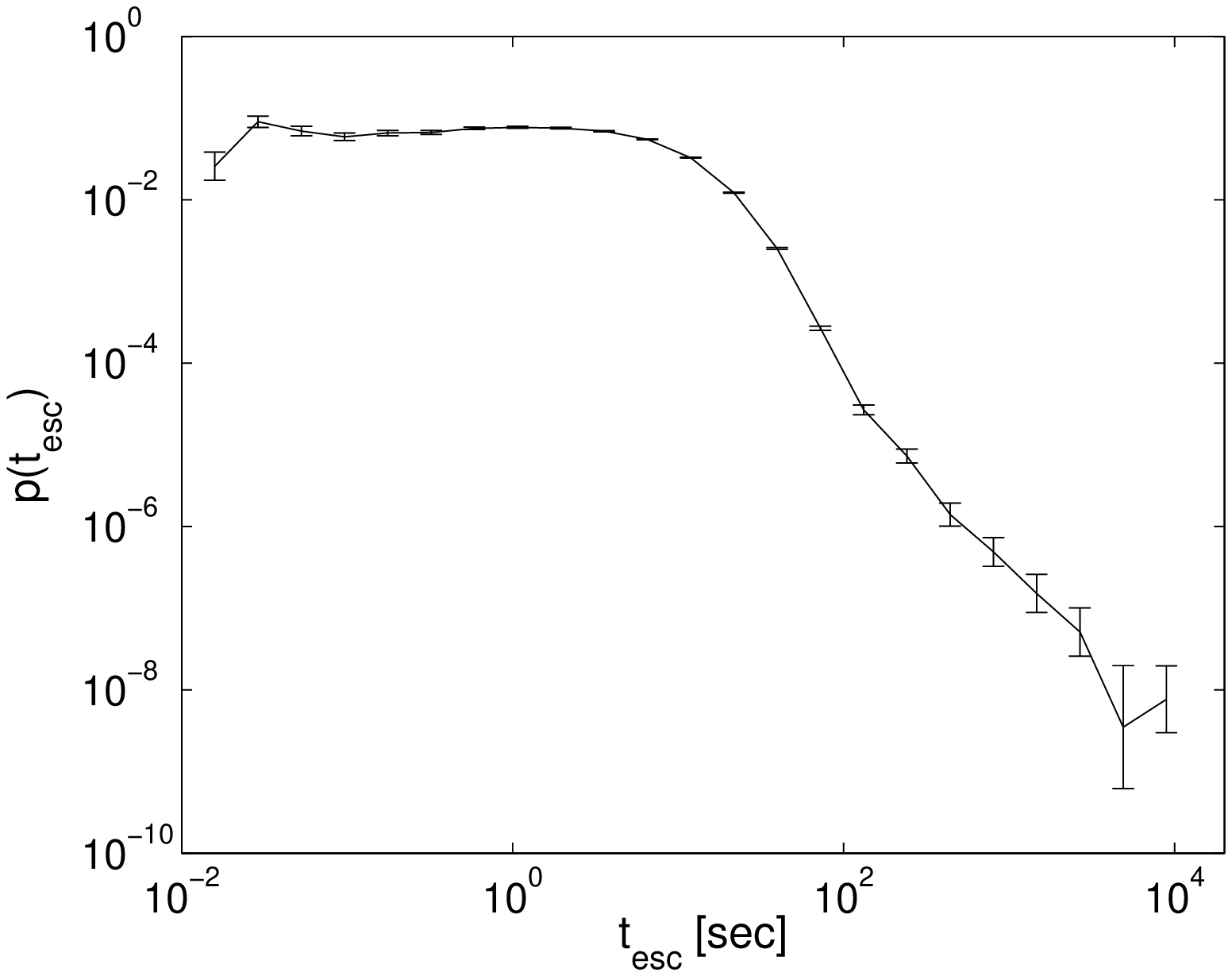}%
        \label{f:F2o:tdistr}}\hfill
    \sidesubfloat[]{\includegraphics[width=0.40\columnwidth]{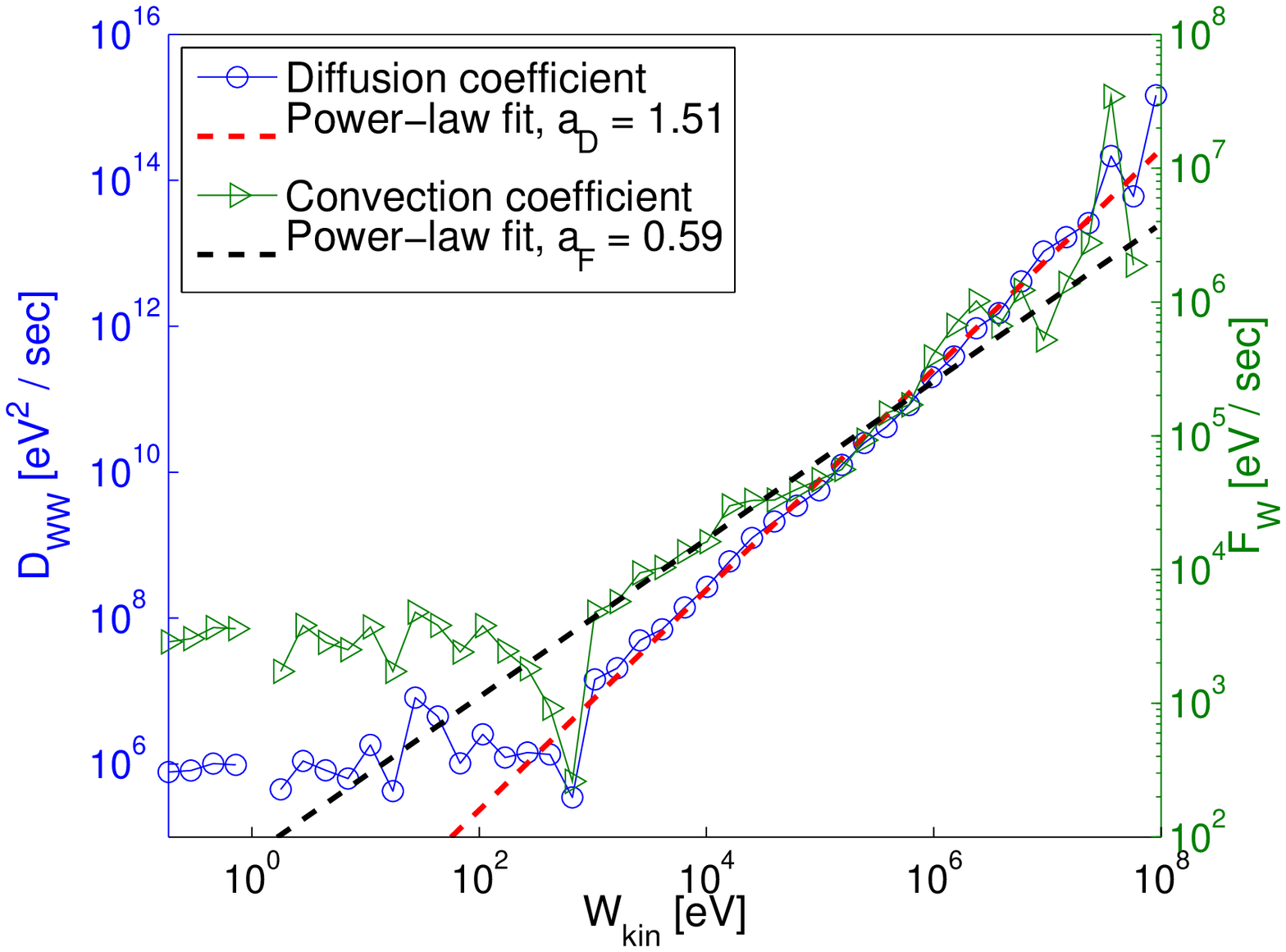}%
        \label{f:F2o:DF_W}}\\
	\sidesubfloat[]{\includegraphics[width=0.50\columnwidth]{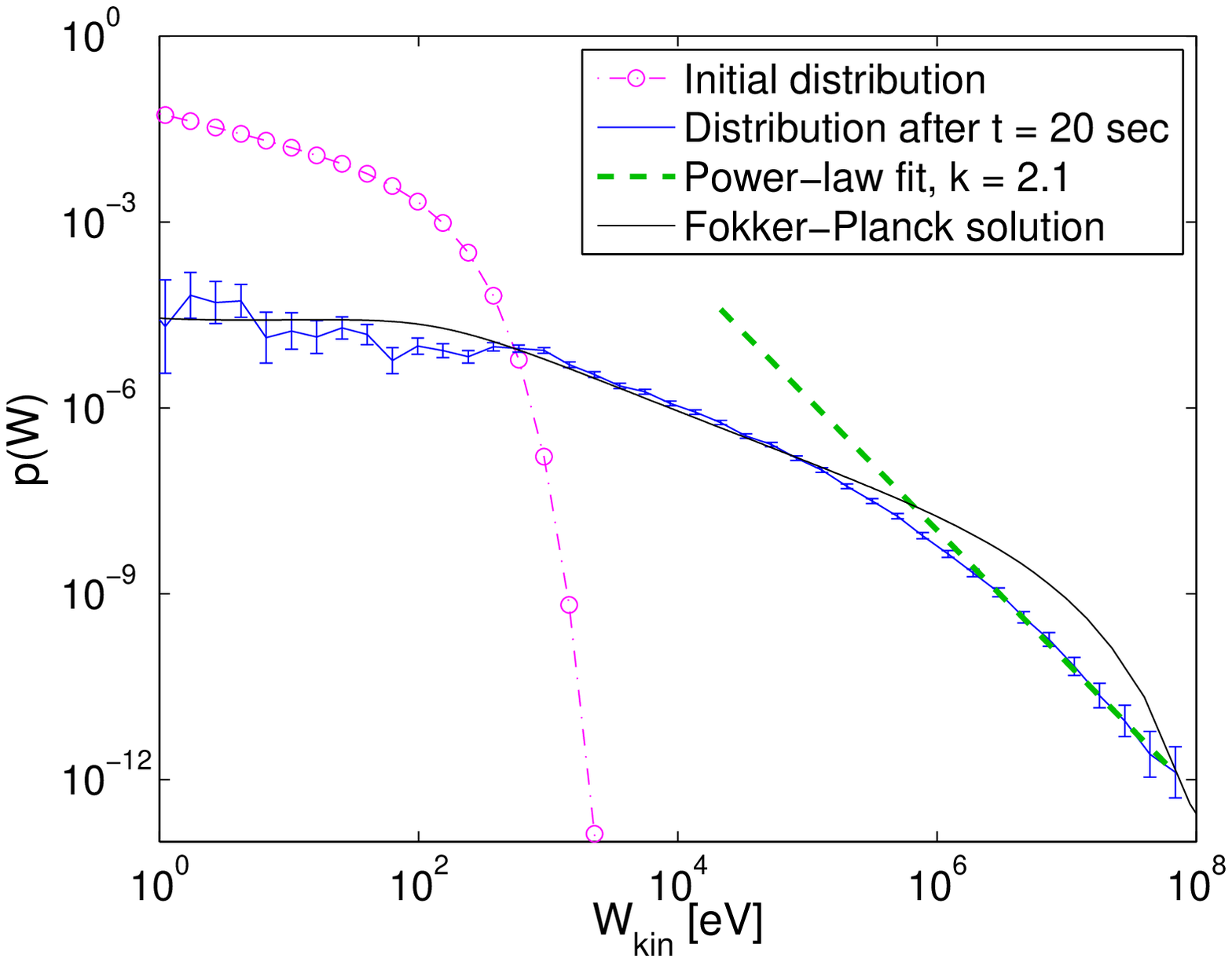}%
        \label{f:F2o:Wdistr}}%
    \caption{\protect\subref{f:F2o:mW} The kinetic energy of the electrons remaining inside the box as a function of time (blue); their mean energy (black) with an exponential fit (green); the kinetic energy of three typical electrons. 
    \protect\subref{f:F2o:mWlog} The 
    transient phase of the 
    temporal evolution of the mean kinetic energy of 
    all the particles (black), 
    %and before it reaches its exponential shape (at $\approx \SI{6}{sec}$),
    together with a power-law fit ($\sim t^{1.6}$, green). \protect\subref{f:F2o:tdistr} The distribution of the escape time of the electrons. \protect\subref{f:F2o:DF_W} The energy diffusion and convection coefficients  as functions of the kinetic energy at time $t = \SI{20}{sec}$. \protect\subref{f:F2o:Wdistr} Energy distribution at $t = 0$ and $t = \SI{20}{sec}$ (stabilized) for the electrons remaining inside the box, and the corresponding solution of the Fokker-Planck equation.}\label{f:F2o}
\end{figure}

The temporal evolution of the mean kinetic energy of the electrons that remain inside the simulation box, by using the parameters stated above, is presented in Fig.~\ref{f:F2o:mW}, along with the kinetic energy evolution of some typical energetic electrons. The energy increases exponentially (after a initial period of a few seconds), as expected from the analysis performed initially by Fermi (see eq.~\ref{Energy}). Using the analytical expression derived by Fermi, eq.~\eqref{rate}, we estimate $t_{\rm acc_{th}} = (3\lambda_{\rm sc} c)/(4V_A^2)\approx \SI{8}{sec}$. We can also estimate the acceleration time from our simulation by making an exponential fit to the asymptotic exponential form of the mean kinetic energy,
%from all electrons,
as predicted by eq.~\eqref{Energy},
%; this fit results in a rough numerical estimation of
which yields an acceleration time of $t_{\rm acc_{num}}\approx \SI{9}{sec}$, a value close to the analytical estimate.
 In Fig. \ref{f:F2o:mWlog}, we show 
%more in detail the transient phase, the brief period before 
the mean energy
% increase starts its exponential increase.  During the transient phase, the
% particles have not yet been accelerated to reach relativistic energies, 
of the particles that remain inside the box during the transient phase
(the first 10 seconds), which
%and their mean energy 
increases as $<W(t)> \sim t^{1.6}$, thus following a scaling close to the prediction of the hard sphere approximation (see Eq.  \ref{meanhs}), and being in correspondence with the functional form of the convection coefficient 
as estimated below. 
%In the subsequent phase, the particles in the tail ($W>1-10 MeV$) dominate the
% energetics.

Fig.~\ref{f:F2o:tdistr} presents the escape time distribution for all electrons (i.e. the time they have reached any boundary); it starts as a uniform distribution at low values and turns over to a power-law distribution at large values. Here we use the median value $(\approx \SI{8}{sec})$ as an estimate for a characteristic escape time from the system, $t_{\rm esc}$.

The energy distribution function of the electrons remaining inside the box is a synthesis of a hot plasma with a mean temperature $\approx \SI{100}{keV}$ and a power-law tail (see Fig.~\ref{f:F2o:Wdistr}). The power-law index is $k \approx \num{2.1}$ after about $\SI{20}{sec}$ and the power-law tail is extended up to $\SI{100}{MeV}$. If we use the values for $t_{\rm acc}$ and $t_{\rm esc}$ estimated above, the simple expression in Eq.~\eqref{index} gives an estimate of the slope of the tail as $k = 1 + t_{\rm acc}/t_{\rm esc} \approx 1 + 9/8 \approx 2.1$, which is identical to the numerical result derived from the simulation.  Using the hard sphere model (see Eq. \ref{heparticles}),  we obtain a very similar result for the power law index 
($k \approx 1.9$). 

\begin{figure}[ht]
    \sidesubfloat[]{\includegraphics[width=0.40\columnwidth]{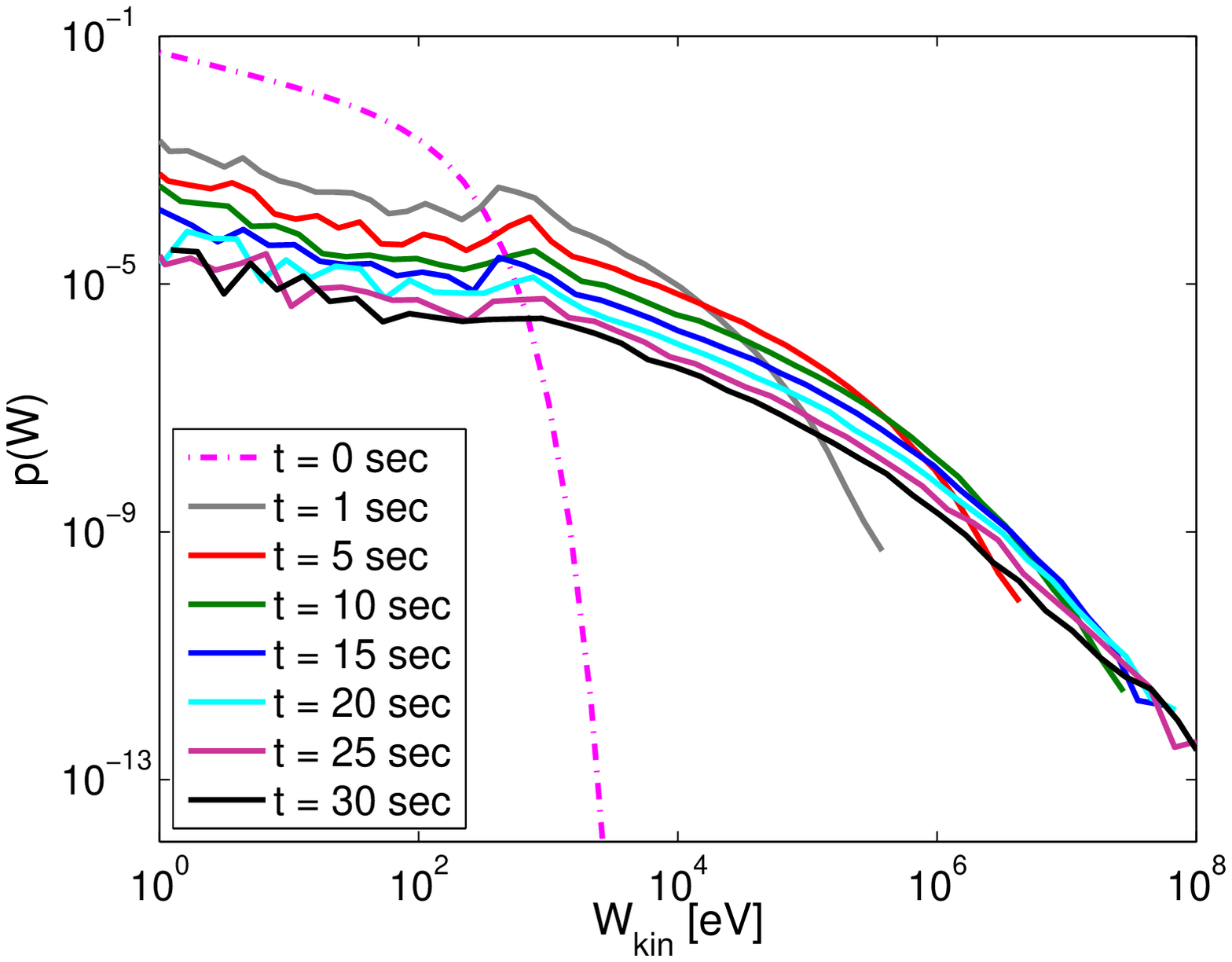}%
        \label{f:F2oTimes:nW}}\hfill
	\sidesubfloat[]{\includegraphics[width=0.40\columnwidth]{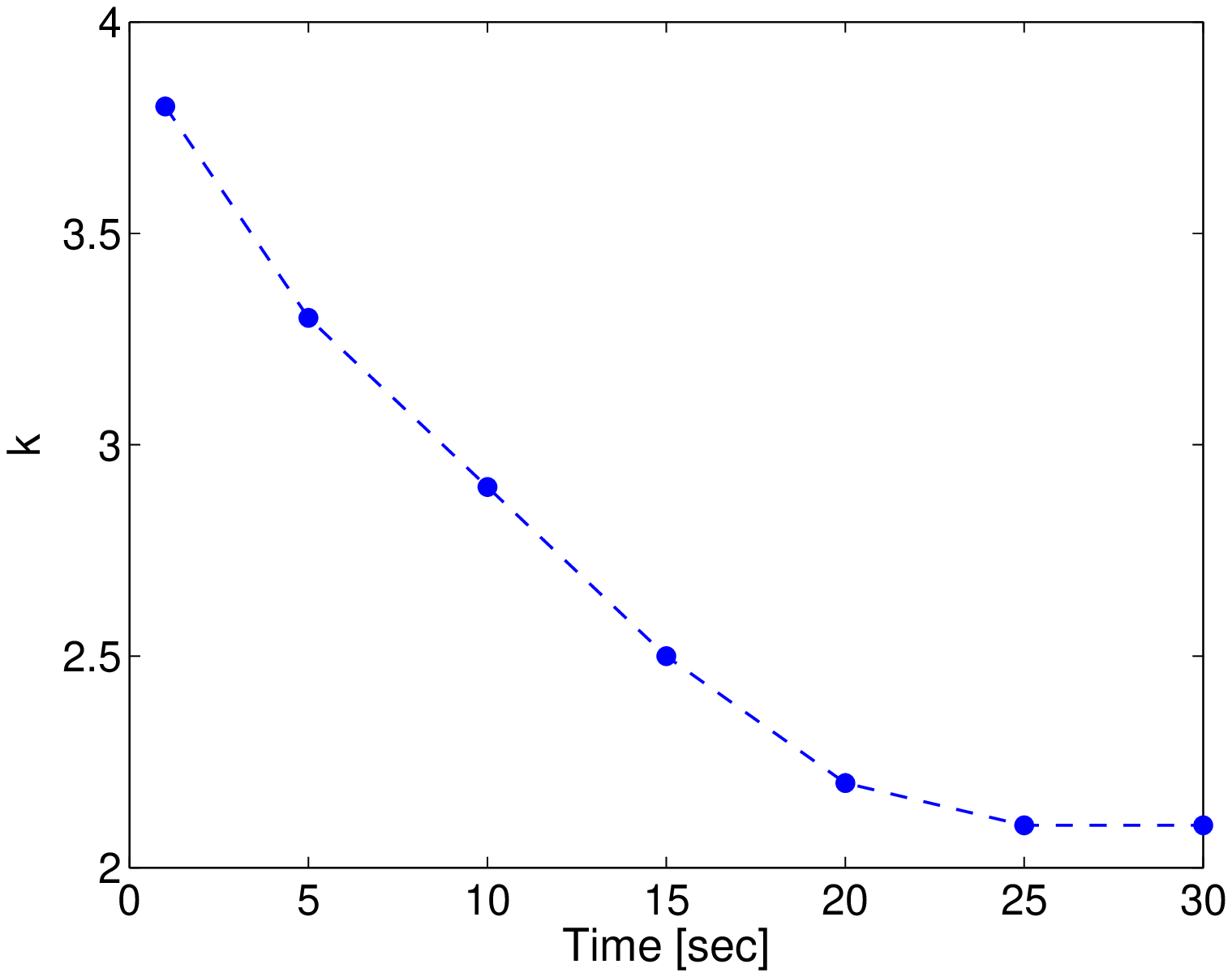}%
        \label{f:F2oTimes:k}}%
    \caption{\protect\subref{f:F2oTimes:nW} Energy distribution at various times for the electrons that remain inside the box. \protect\subref{f:F2oTimes:k} The power-law index of the tail of the energy distribution at various times.}\label{f:F2oTimes}
\end{figure}

%Another characteristic of the energy distribution is the peak of the distribution located near $W = \SI{500}{eV}$. This is the result of the first term in eq.~\eqref{energyF}, which is always positive; for the present setup $2\gamma_{\rm cl}(W_{\rm th}+W_{\rm rest})\frac{V^2}{c^2} \approx \SI{540}{eV}$, where $W_{\rm th}$ is the thermal energy of the particle and $\gamma_{\rm cl}$ is the Lorentz factor for the cloud.

In Fig.~\ref{f:F2o:DF_W}, the diffusion and convection coefficients at $ t = \SI{20}{sec}$, as functions of the energy, are presented. The estimate of the coefficients is based on eqs.~\eqref{eq:DWW} and \eqref{eq:FW}, with $\Delta t$ small, whereto we monitor the energy of the electrons at a number of regularly spaced monitoring times $t^{(M)}_k$, $k=0,1,\ldots,K$, with $K$ typically chosen as $200$, and we use $t=t^{(M)}_{K-1}$, $\Delta t = t^{(M)}_{K} - t^{(M)}_{K-1}$ in the estimates. Also, in order to account for the conditional averaging in eqs.~\eqref{eq:DWW} and \eqref{eq:FW}, we divide the energies $W\left(t^{(M)}_{K-1}\right)_i$ of the particles into a number of logarithmically equi-spaced bins and perform the requested averages separately for the particles in each bin. As Fig.~\ref{f:F2o:DF_W} shows, both transport coefficients exhibit a power-law shape for energies above \SI{1}{keV}, $D(W) = 223.57\ W^{1.51}$ and $F(W) = 73.47\ W^{0.59}$.  {\bf These results agree quite well with the estimates reported above (Eqs. \ref{conParker} and \ref{diffParker}) from the hard sphere model of \cite{Parker58, Ramaty79}}.

In order to verify the estimates of the transport coefficients, we insert them into the FP equation 
(eq.~\ref{diff}) 
and solve the FP equation numerically in the form
\begin{equation} \label{diff2}
\frac{\partial n}{\partial t} +
\frac{\partial}{\partial W} \left[F n -\frac{\partial (D n) }{\partial W} \right] =
-\frac{n}{t_{\rm esc}} ,
\end{equation}
including the escape term $-n/t_{\rm esc}$, with $ t_{\rm esc} = 8\,$s
the median value from Fig.~\ref{f:F2o:tdistr}.
The transport coefficients are inserted in the form of the fit above 1 keV and set to constant below 1keV such that they are continuous at the transition. 
For the integration of the FP equation on the semi-infinite energy interval $[0,\infty)$, we use the pseudospectral method, based on the expansion in terms of rational Chebyshev polynomials in energy space, combined with the implicit backward Euler method for the time-stepping \citep[see e.g.][]{Boyd2001}. The resulting energy distribution at final time is also shown in Fig.~\ref{f:F2o:Wdistr}, and it turns out to coincide very well with the distribution from the electron simulation in the intermediate energy range that corresponds to the heating of the population, the power-law tail can though not be reproduced in shape by the FP solution.% The differences at low energies are of less importance and can be attributed to the different form of the initial conditions, a 3D Maxwellian for the FP equation, and a 1D Maxwellian for the test-particles.

The heating and the power-law tail in the energy distribution function are formed almost from the start of our simulation and reach their asymptotic shape on a time scale comparable with the acceleration time (see Fig.~\ref{f:F2oTimes:nW}). In the beginning, the power-law index is $\approx 5$ and it gradually decreases until it reaches an asymptotic value \num{2.1} in about \SI{20}{sec} (almost twice the acceleration time), as shown in Fig.~\ref{f:F2oTimes:k}.

\begin{figure}[ht]
	\includegraphics[width=0.45\columnwidth]{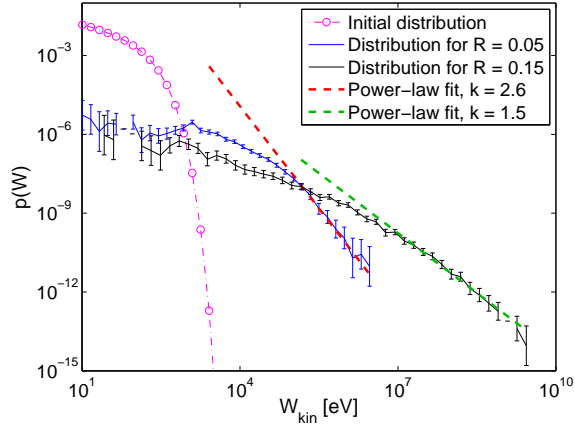}%
	%\sidesubfloat[]{\includegraphics[width=0.45\columnwidth]{figures/nWfit_F2oV53_t35}%
        %\label{f:F2o_V}}%
    \caption{The stabilized energy distribution for electrons remaining inside the box at $t = 0$ and $t = \SI{25}{sec}$  for $R = 0.05 \;\;(\lambda_{\rm sc} = \SI{3.3e8}{cm})$ (blue), and  $R = 0.15\  (\lambda_{\rm sc} = \SI{1.1e8}{cm})$ (black).}\label{f:F2o_R}
    %\protect\subref{f:F2o_V} Energy distribution for particles remaining inside the box at $t = 0$ and $t = \SI{35}{sec}$ when the velocity of the cloud follows a power-law distribution with index $5/3$, and values in $[0.1V_A,\ V_A]$.}
\end{figure}

In the case of parameters we considered, the escape time  is locked to the acceleration time ($t_{\rm acc} \approx t_{\rm esc}$), therefore the density of the scatterers, which controls the mean free path, is the most important parameter for our system. A parametric study of the evolution of the energy distribution of the particles, as we vary the density of the scatterers $0.05 < R < 0.15$ (i.e.~$\SI{3.3e8}{cm} < \lambda_{\rm sc} < \SI{1.1e8}{cm}$), keeping the characteristic length of the acceleration volume constant, was made and we find that the escape time varies between $\SI{5}{sec} < t_{\rm esc} < \SI{8}{sec}$, while the acceleration time decreases from $\approx \SI{8}{sec}$ to $\approx \SI{4}{sec}$. The power-law tail index also decreases and it remains close to $3 \gtrapprox k \gtrapprox 1.5$. In Fig.~\ref{f:F2o_R} we show stabilised distribution functions for $R=0.05$ and $R=1.5$.  The time when the $k$-index stabilizes varies between \num{20} to \SI{25}{sec}.

%\begin{figure}[htb]
%	\includegraphics[width=0.70\columnwidth]{figures/nWfit_F2o_R05R15_t25}%
%	\caption{Energy distribution for particles remaining inside the box at $t = 0$ and $t = \SI{25}{sec}$ (stabilized) for $R = 0.05$ and $R = 0.15$.}\label{f:F2o_R}
%\end{figure}

We have preformed an extensive analysis of the role of the free parameters 
($R$, as just described, and also $N$ and $L$) in our results and we conclude that, if in our setup for the physical system under investigation  we keep the mean free path of the particles ($\lambda_{\rm sc}$) constant, the main results presented above remain the same.
%\bnote{\textbullet~[Completely omit non-constant $V$?[}
  %The last important parameter in the setup of the model is the velocity %of the scatterer $V$ in eq.~\eqref{energyF}. So far we consider a %cloud with the Alfv\'en velocity $V = V_A$. We have also tried a %power-law distribution for $V$ with values in $[0.1V_A,\ V_A]$ and %various indices (5/3, 2, 5/2). This change does not affect the shape %of the energy distribution, and as expected from eq.~\eqref{rate}, %the acceleration time increases and the energy gain decreases. %Consequently, the escape time increases but not as much as the %acceleration time, we thus expect a $k$-value greater than the $V = %V_A$ case; this agrees with the simulation results (see Fig.~%\ref{f:F2o_V} for an example), however the particles escape before the %$k$-index saturates.

\begin{figure}[ht]
    \sidesubfloat[]{\includegraphics[width=0.40\columnwidth]{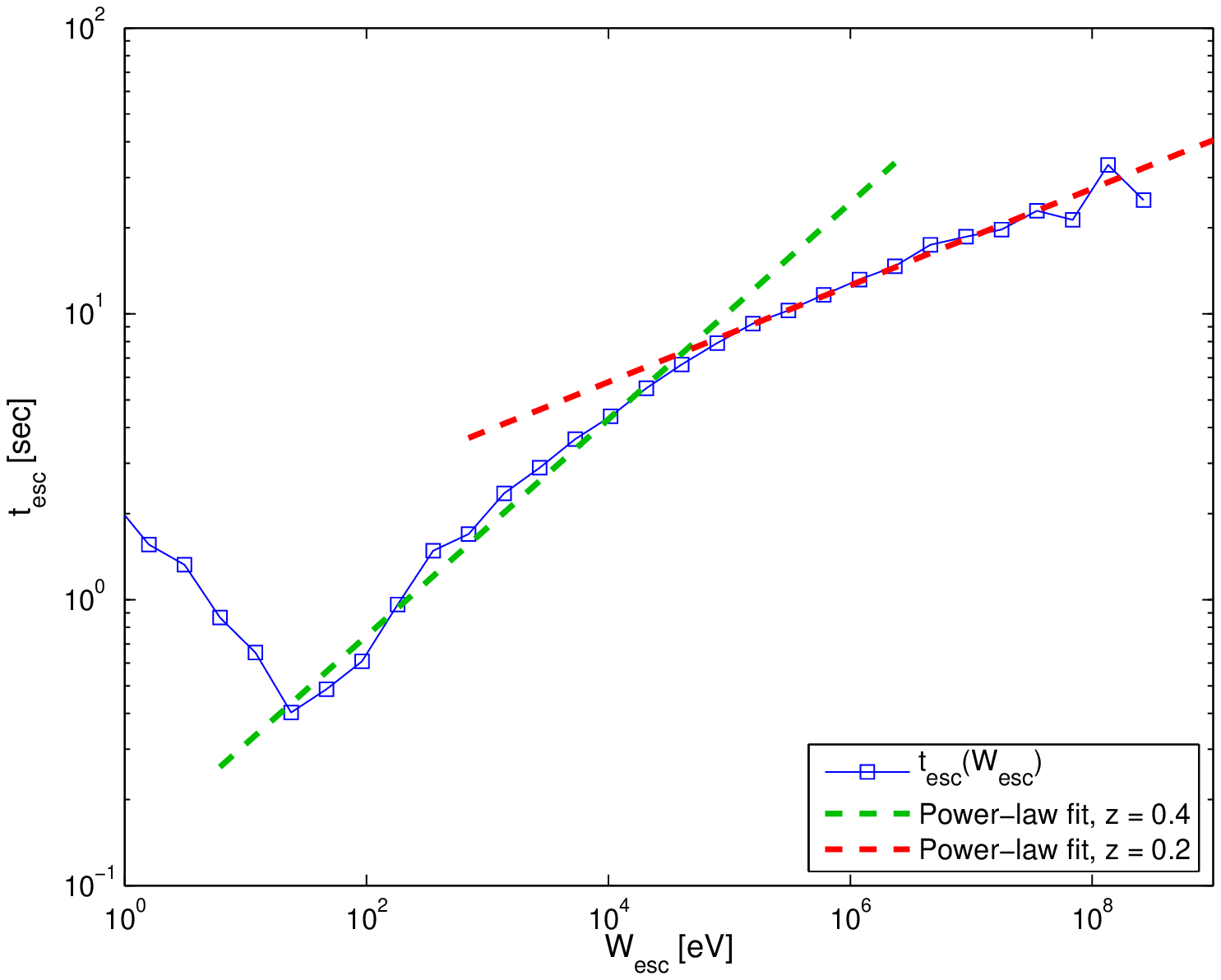}%
        \label{f:F2o_escDistr:tW}}\hfill
	\sidesubfloat[]{\includegraphics[width=0.40\columnwidth]{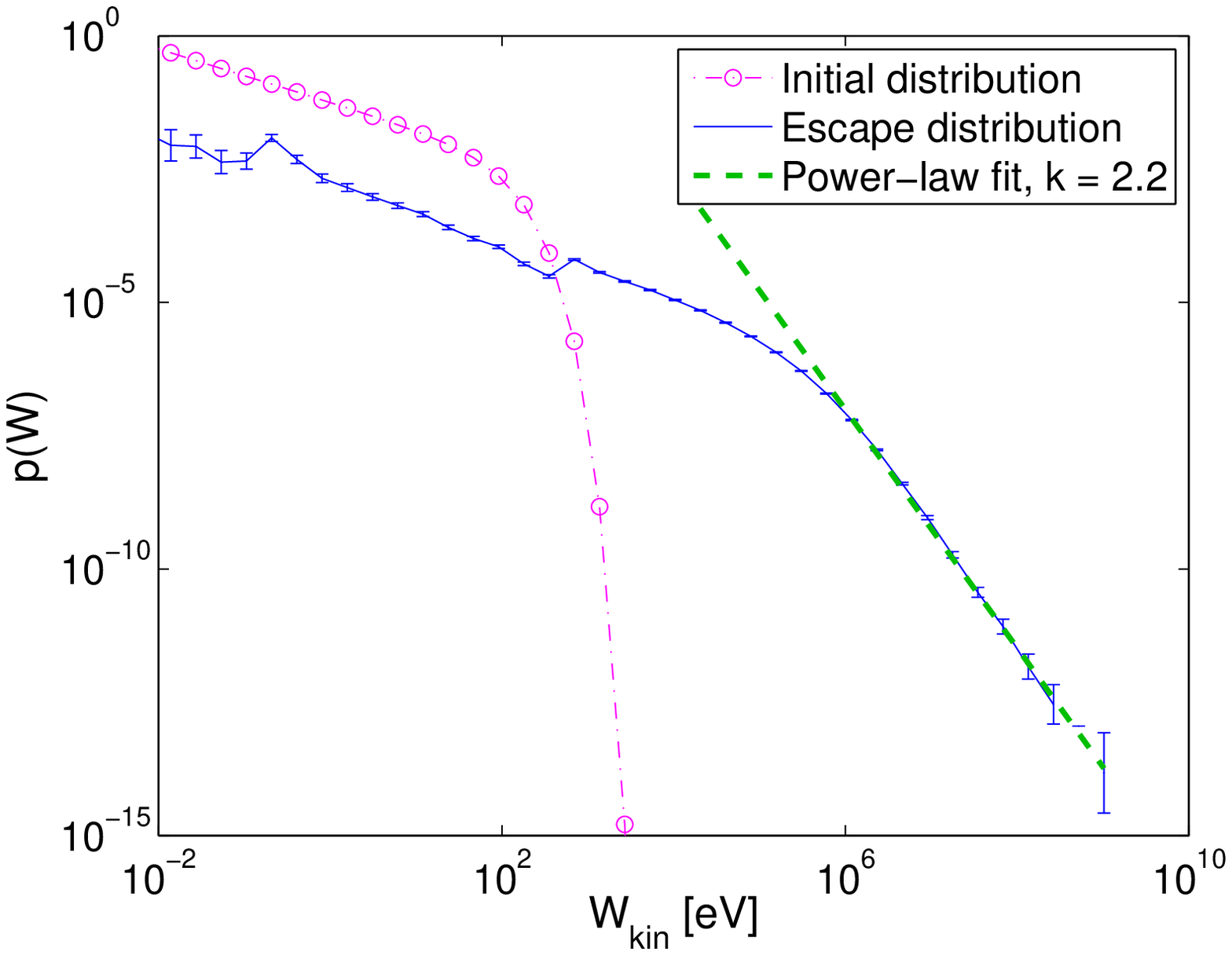}%
        \label{f:F2o_escDistr:nW}}\\%
    \sidesubfloat[]{\includegraphics[width=0.40\columnwidth]{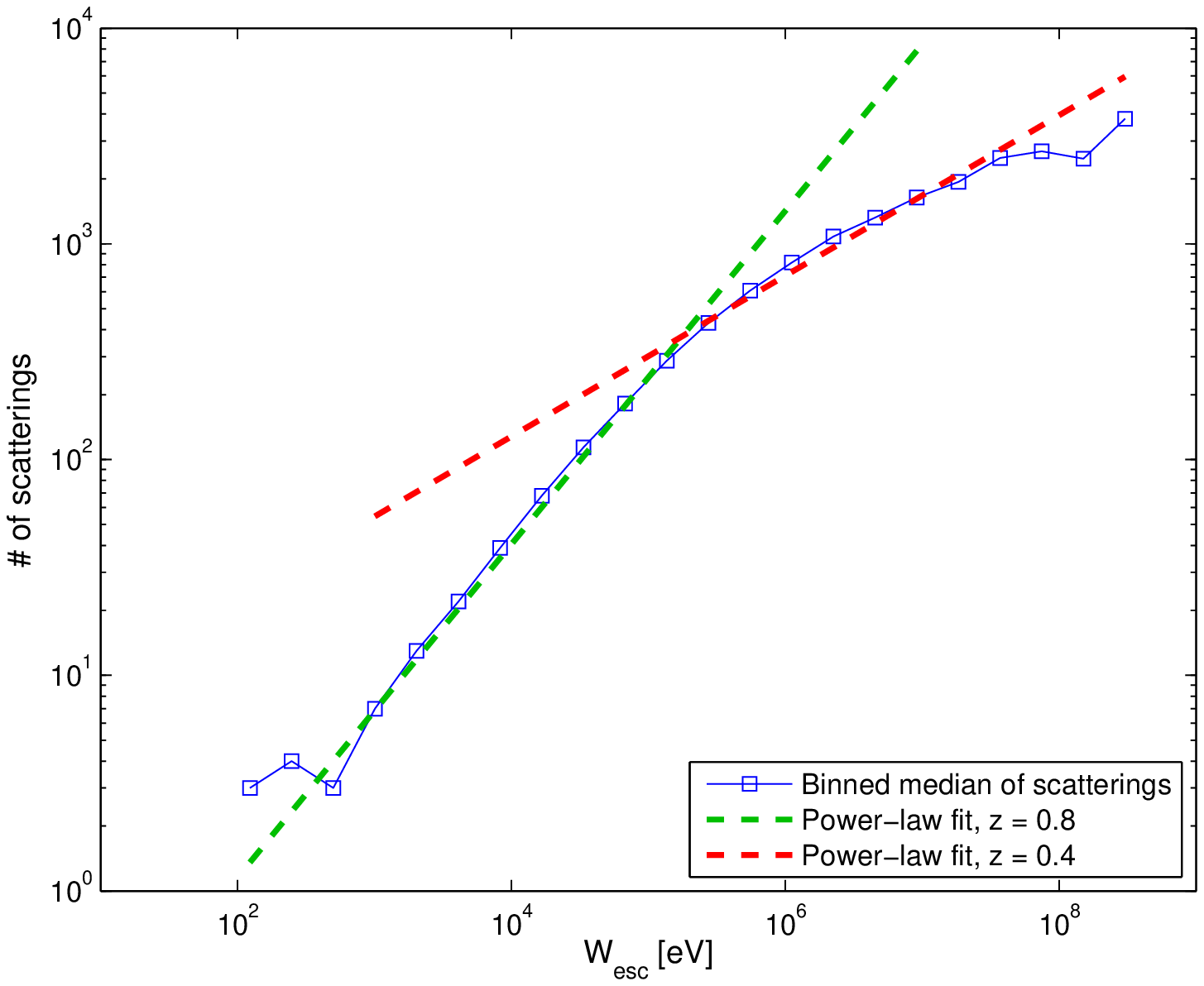}%
        \label{f:F2o_escDistr:kicks}}%
    \caption{\protect\subref{f:F2o_escDistr:tW} The escape time as a function of the escape energy of the electrons. \protect\subref{f:F2o_escDistr:nW} The energy distribution of the electrons that have escaped from the box. \protect\subref{f:F2o_escDistr:kicks} The mean number of scatterings per particle as a function of the escape energy.}
        \label{f:F2o_escDistr}
\end{figure}

Depending on the initial energy and the energization in the scattering events, each of the escaping electrons leaves the acceleration volume with a different energy and at a different time. We have already shown the distribution of the escape times $t_{\rm esc}$ in Fig.~\ref{f:F2o:tdistr}. In Fig.~\ref{f:F2o_escDistr:nW}, we present the energy distribution of the escaped electrons, $p(W_{\rm esc})$, which has a shape very similar to the one of the energy distribution shown in Fig.~\ref{f:F2o:Wdistr} of the particles that remain inside the simulation box until saturation, and it exhibits a thermal distribution for the low energy particles $(W<100 KeV)$ and a power-law tail for the relativistic particles with energies $W>1MeV.$ In Fig.~\ref{f:F2o_escDistr:tW}, we show the mean escape time as a function of the energy with which the particles escape from the turbulent volume, and it follows a power-law distribution, $t_{\rm esc} \propto W_{\rm esc}^z.$ We observe two distinct regions: for the non-relativistic energies, \SI{10}{eV}--\SI{.e5}{eV}, the exponent is $z = 0.4$, while for the relativistic particles, \SI{.e5}{eV}--\SI{.e9}{eV}, the exponent drops to $z = 0.2.$ The relation of  $t_{esc}$ with the escape energy of the electrons can be understood by the fact that energization should closely be connected with the trapping  of the particles  by the ASs.  The correlation of the trapping of the electrons with their final energy can be demonstrated when considering the mean number of scatterings the escaping particles suffer before they escape from the simulation box as a function of their energy, as shown in Fig.~\ref{f:F2o_escDistr:kicks}, from which it is clear that the energization depends  strongly on the trapping of the particles inside the turbulent volume, and thus, the longer they stay inside the more they are accelerated.  This result is clearly very different from the $t_{\rm esc}$ used in the literature on stochastic weak turbulence acceleration \cite[$t_{\rm esc} \approx L/v$; see][]{Miller90, Petrosian12}, but it seems to agree qualitatively with the observational results obtained by \cite{Petrosian10}, although their estimate is based on a very simple magnetic topology, e.g.\ a simple magnetic loop. The trapping of the energetic particles inside the turbulent volume, where acceleration and heating takes place, seems to solve another well documented observation, the relatively long life of Hard X-ray sources following CMEs in the high corona \citep{Krucker07}. As we have shown, the strong trapping of the electrons for tens of seconds inside the ASs can explain this observation.

\begin{figure}[ht]
	\sidesubfloat[]{\includegraphics[width=0.40\columnwidth]{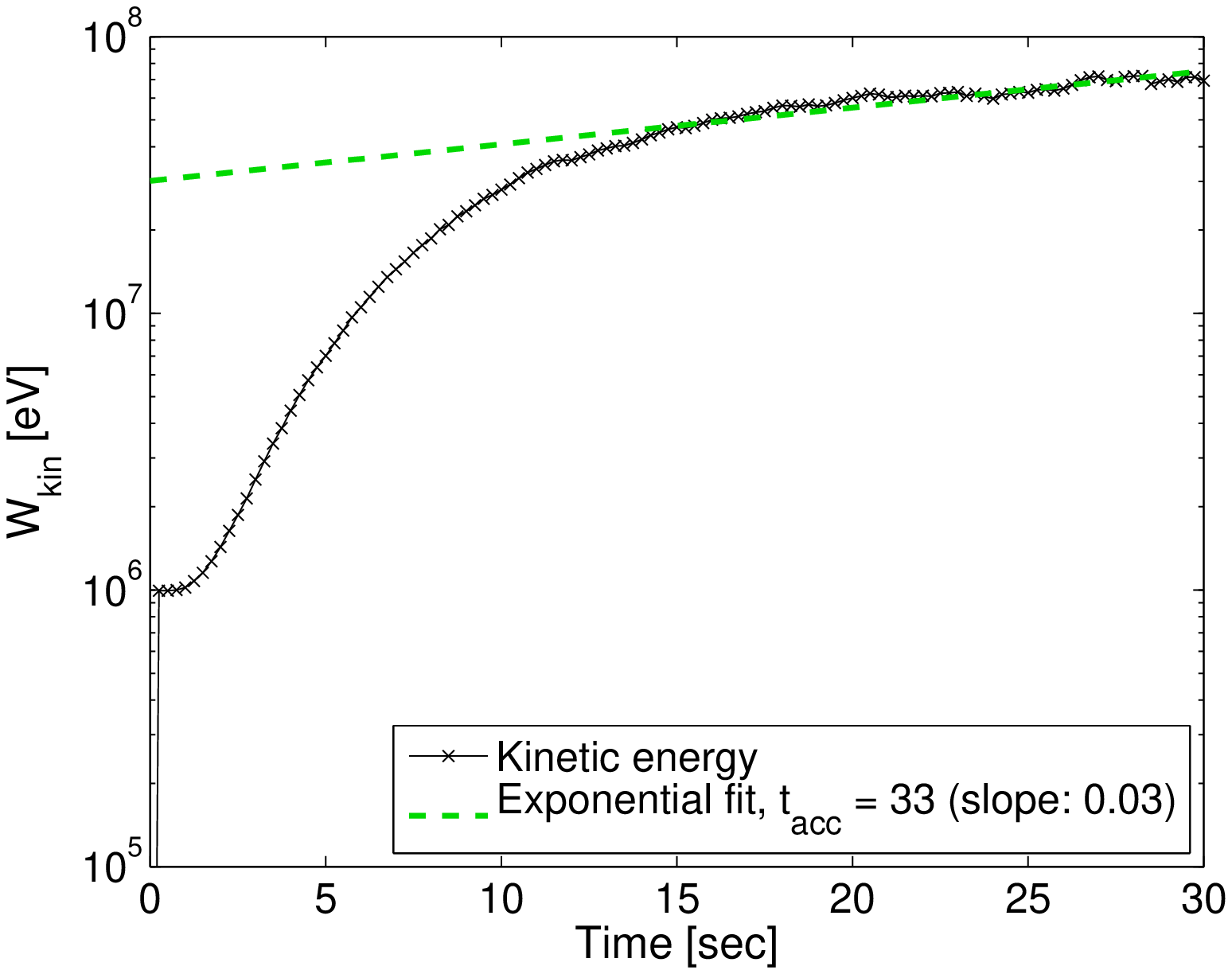}%
        \label{f:F2oI:mW}}\hfill
    \sidesubfloat[]{\includegraphics[width=0.40\columnwidth]{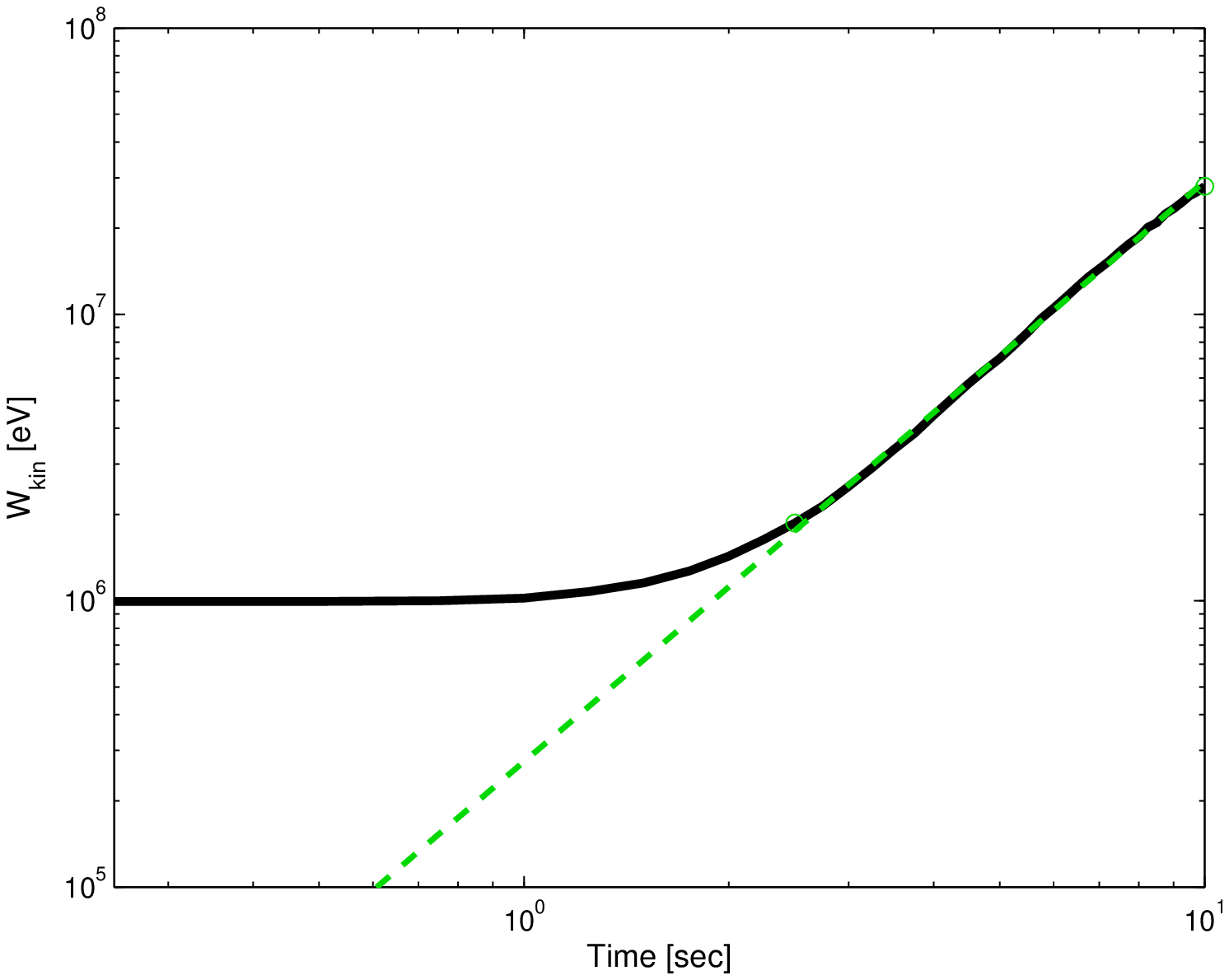}%
        \label{f:F2oI:mWlog}}\\
	\sidesubfloat[]{\includegraphics[width=0.40\columnwidth]{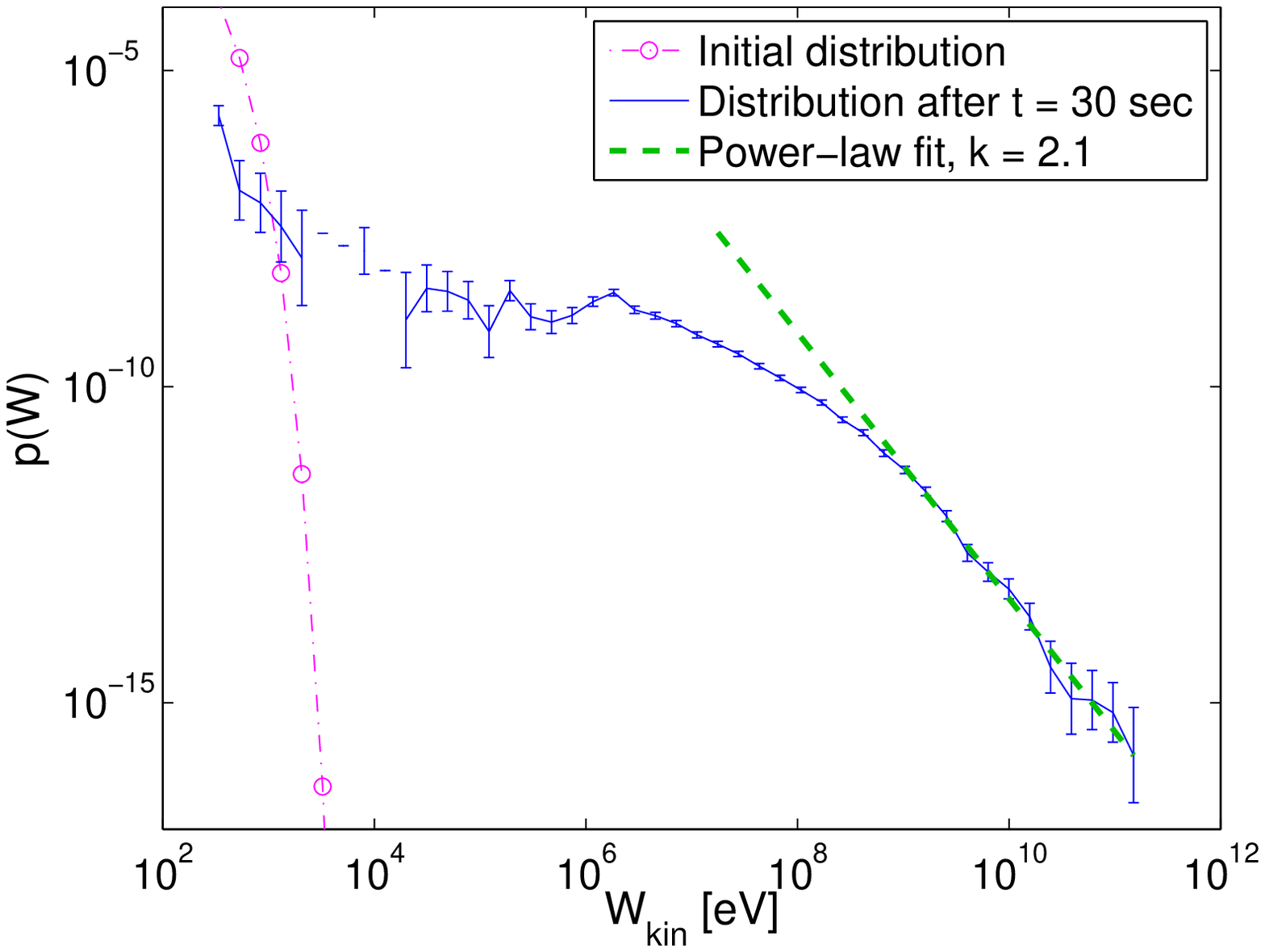}%
        \label{f:F2oI:Wdistr}}%
    \caption{\protect\subref{f:F2oI:mW} The mean kinetic energy of the ions remaining inside the box, as a function of time (black) with an exponential fit (green). 
    \protect\subref{f:F2oI:mWlog} The  
    temporal evolution of the mean kinetic energy of 
    all the ions in the transient phase (the first seconds, black), along with a power-law fit ($\sim t^2$, green).
    %; the kinetic energy of three typical particles. \protect\subref{f:F2o:tdistr} The distribution of the escape time of the particles. \protect\subref{f:F2o:DF_W} The energy diffusion and convection coefficients (normalized for comparison) as functions of the kinetic energy.
    \protect\subref{f:F2oI:Wdistr} Energy distribution at $t = 0$ and $t = \SI{30}{sec}$ (stabilized) for the ions remaining inside the box.}\label{f:F2oI}
\end{figure}

Our results so far refer to electrons. The ions, in the asymptotic state, do not have significant differences from the evolution of the electrons, other than the time scale needed to reach this state. The energy distribution exhibits the same characteristics (heating and  acceleration) and it saturates after \SI{27}{sec}; in Fig.~\ref{f:F2oI:Wdistr} we show the saturated distribution of the ions remaining inside the domain at $t = \SI{30}{sec}$, it extends up to tens of GeV. The median escape time is $\approx \SI{26}{sec}$, and the acceleration time calculated from the analytical expression in Eq.~\eqref{rate} remains unchanged; the numerically estimated value, through eq.~\eqref{Energy}, is $t_{\rm acc_{num}} \approx \SI{33}{sec}$ (see Fig.~\ref{f:F2oI:mW};
as for the electrons, the mean energy evolves as a power-law in the the transient phase, see Fig.\ \ref{f:F2oI:mWlog}). Applying these two values in Eq.~\eqref{index} leads to $k \approx 2.1$, which agrees with the measured slope of the power-law tail in the energy distribution.

We can then conclude that SFE can heat and accelerate both ions and electrons in the solar corona to a level as it is observed.
%-----------------------------------------------------------------------------------
\subsubsection{The role of collisions in the evolution of the energy distribution}\label{collisions}
We apply a series modification of the simplified model of \cite{Lenard58}  for the Coulomb collisions of charged particles with a background plasma population of temperature $T$ (which coincides with the colliding particles' initial temperature in our approach),
%{\bf
%\begin{gather*}
%    \frac{\d s}{\d t} = v, \label{e:loss1}\\
%    \frac{\d v}{\d t} =- \nu_{\rm th} v + \sqrt{2 \nu_{\rm th} k_B T/m} \ W_t,
%\label{e:loss}
%\end{gather*}
%}

\begin{equation}\label{e:loss}
\begin{gathered}
    \frac{\d s}{\d t} = v, \\
    \frac{\d v}{\d t} =- \nu_v v + \sqrt{2 \nu_v k_B T/m} \ N_t,
\end{gathered}
\end{equation}
where $k_B$ is the Boltzmann constant, and $N_t$ an independent and identically distributed Gaussian random variable with mean value zero and variance the integration time-step $\Delta t$. Within this model, a particle travels a distance $s$ between two consecutive scatterers (which is an integer multiple of $l$) for a time interval $\tau$, with non-constant velocity due to collisions with the background plasma. Following \cite{Gillespie1996}, we directly use the analytical solution to eq.~\eqref{e:loss},
\begin{equation}\label{e:loss_sol}
    v(t+\tau) = v(t)\mu + \sigma_v N_1,
\end{equation}
with $\mu = e^{-\nu_v\tau}$, $\sigma_v^2 = \frac{k_B T}{m}\left(1-\mu^2\right)$, and $N_1$ a Gaussian random variable with mean 0 and standard deviation 1, which allows to use arbitrarily large time-steps and thus is computationally favorable.
The use of the analytical solution allows to make the \cite{Lenard58} collision model more realistic than it is in its standard form, namely by making the collision frequency velocity dependent, $\nu_v \propto 1/v^3$, as appropriate for a fully ionized plasma \citep[see e.g.][]{Karney86},
with $v$ the instantaneous velocity of a particle before each of the collisionally affected travel events in between subsequent scatterings. 
We also note that $v$ here is the non-relativistic speed that is bounded by the speed of light, the characteristic scaling thus breaks down for very energetic particles.

\begin{figure}[ht]
    \sidesubfloat[]{\includegraphics[width=0.40\columnwidth]{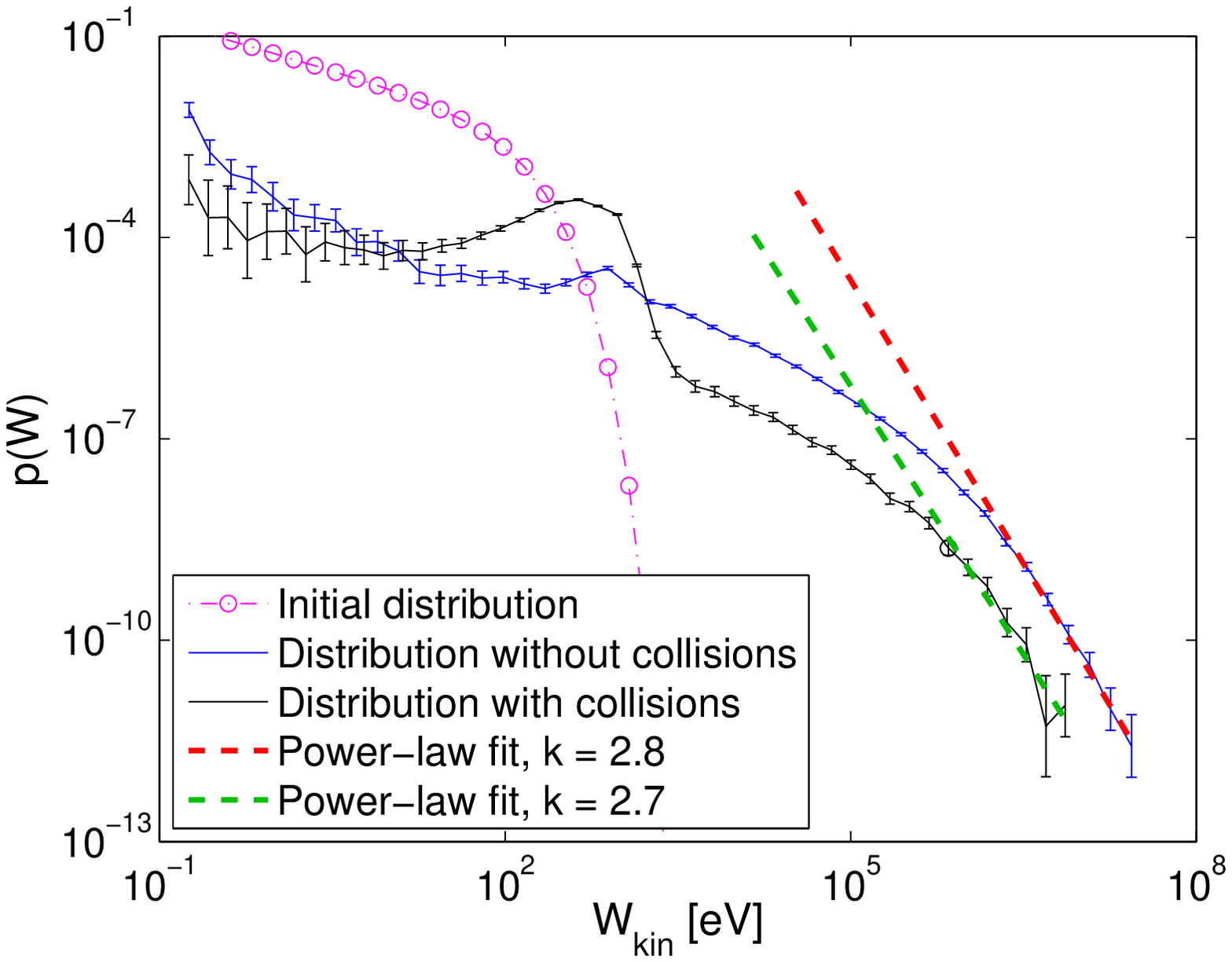}%
        \label{f:F2oC_Wdistr:t10}}\hfill
	\sidesubfloat[]{\includegraphics[width=0.40\columnwidth]{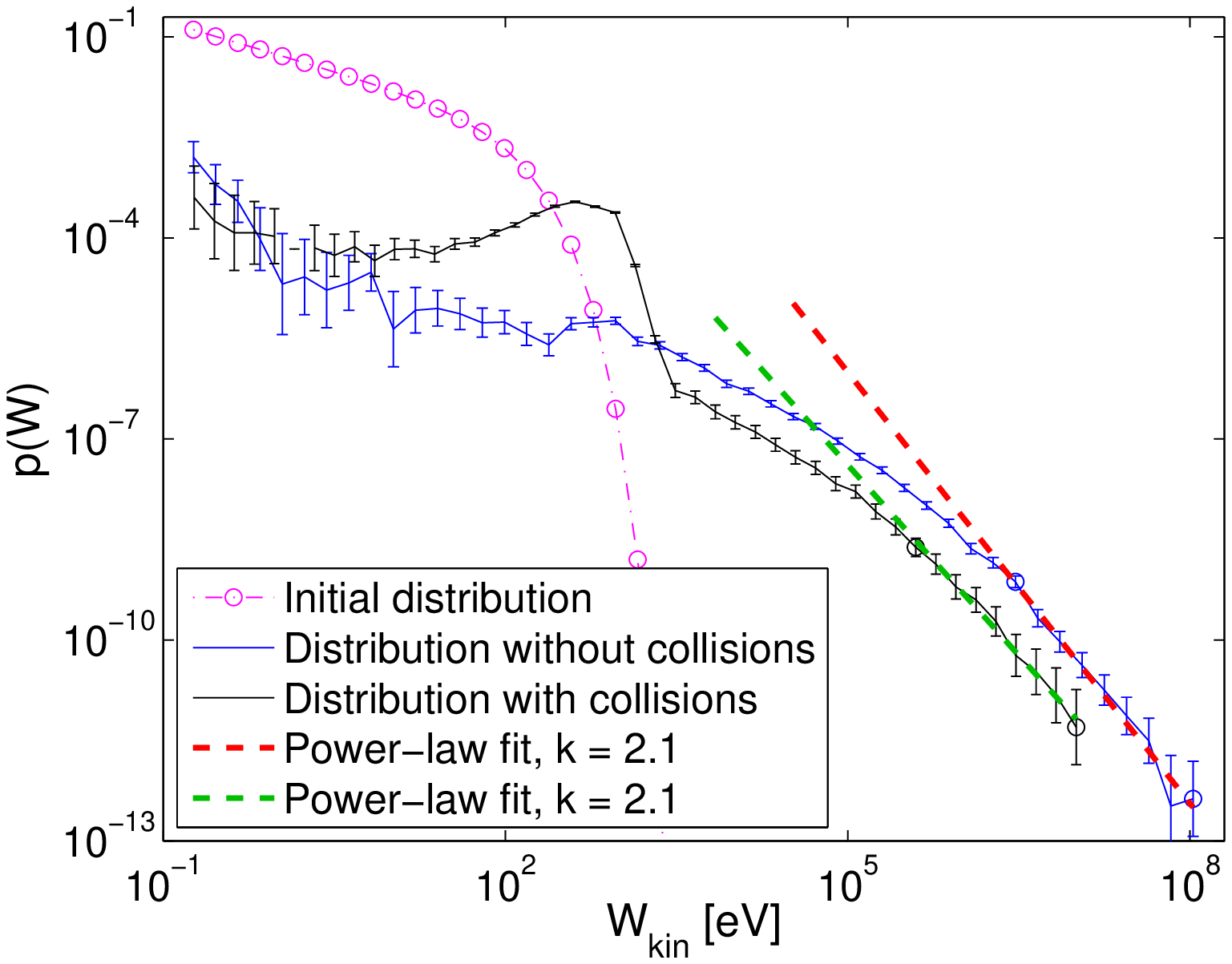}%
        \label{f:F2oC_Wdistr:t25}}%
    \caption{\protect\subref{f:F2oC_Wdistr:t10} Energy distribution for electrons remaining inside the simulation box (a) at $t = 0$ (magenta)   and at $t = \SI{10}{sec}$ with collisions (black)  and  without (blue), and  \protect\subref{f:F2oC_Wdistr:t25} at $t = \SI{25}{sec}$, after the stabilization.}\label{f:F2oC_Wdistr}
\end{figure}

In Fig.~\ref{f:F2oC_Wdistr}, we present the evolution of the energy distribution of electrons  with and without collisions at two instances, one before ($t = \SI{10}{sec}$), and one after its stabilization ($t = \SI{25}{sec}$). Even though the mean free path for the collisions is almost 20 times smaller than the mean free path for the scattering ($\lambda_{\rm coll} = \SI{7.67e6}{cm}$), for the characteristic time scale of the interaction we consider (e.g.~the acceleration time), the distribution separates into the low to intermediate energy part (\SI{300}{eV} -- \SI{3}{keV}), where the collisions dominate and the evolution of the energy distribution approaches a Maxwellian distribution, and the higher energy part, where the energization dominates and the electrons are accelerated, forming a power-law tail with index similar to the one of the collisionless case, though slightly reduced in extent.

%-----------------------------------------------------------------------------------
\subsection{Periodic boundary conditions}\label{periodic}
\begin{figure}[!ht]
    \sidesubfloat[]{\includegraphics[width=0.40\columnwidth]{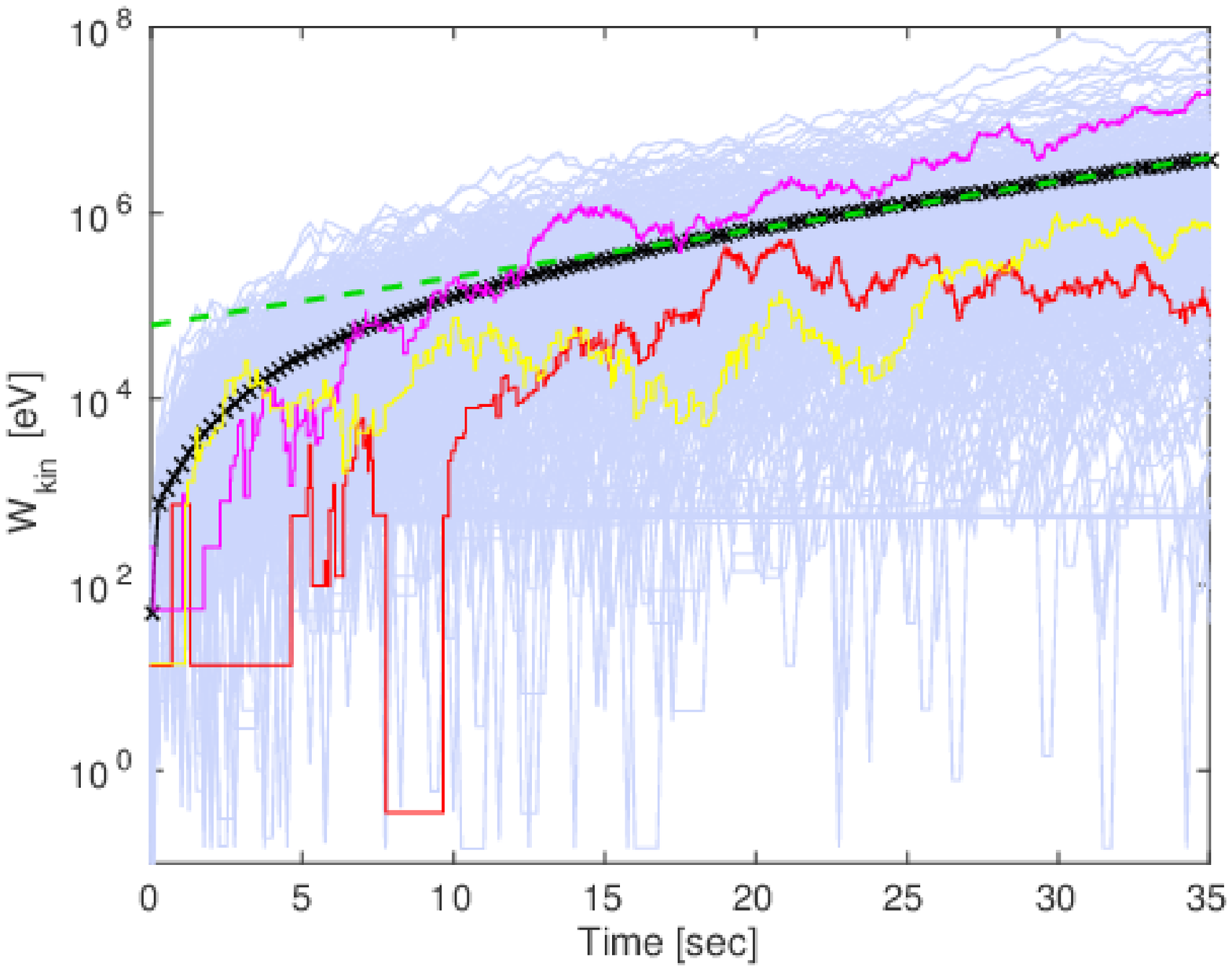}%
		\label{f:F2p:mW}}\hfill
	\sidesubfloat[]{\includegraphics[width=0.40\columnwidth]{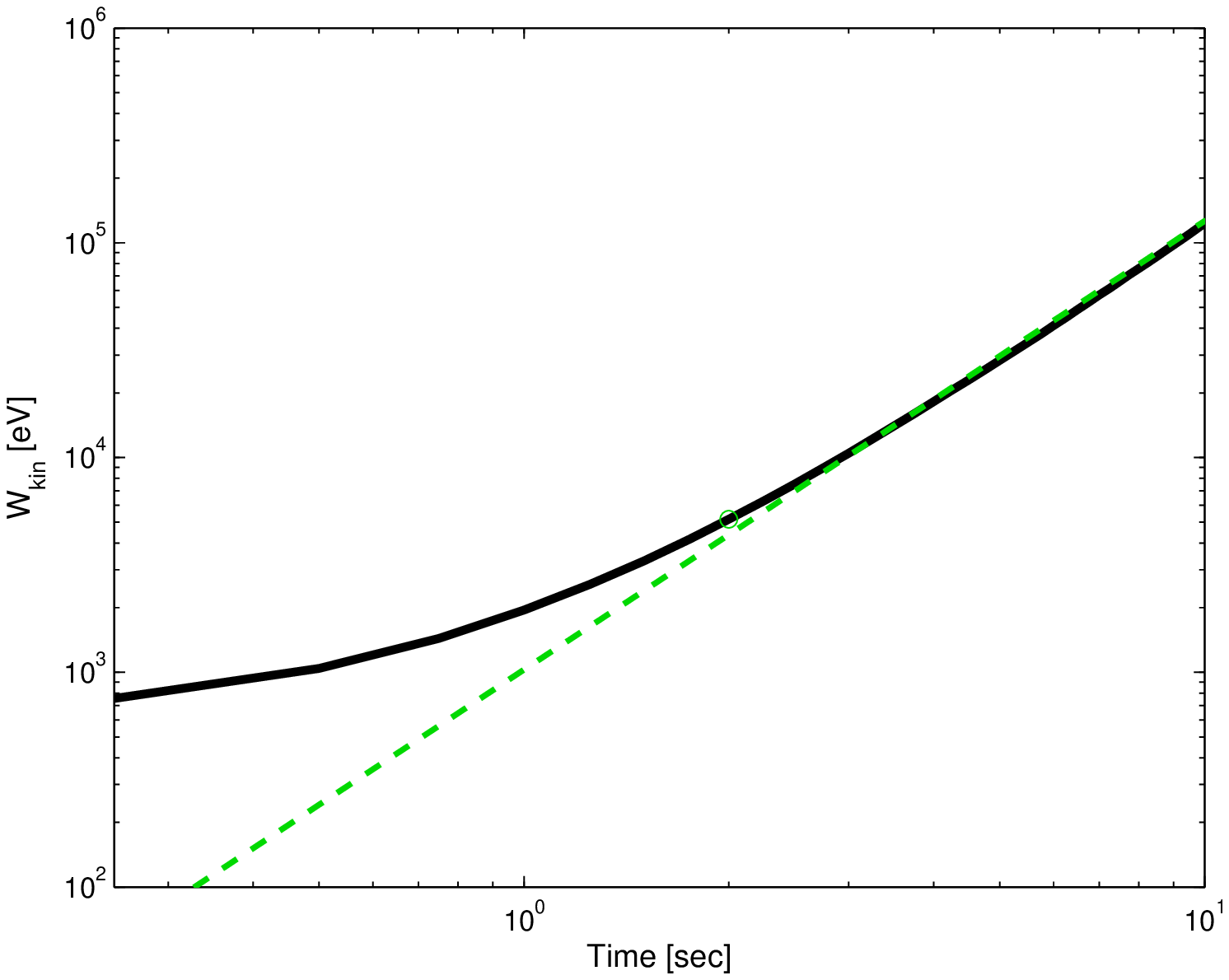}%
        \label{f:F2p:mWlog}}\\
	\sidesubfloat[]{\includegraphics[width=0.40\columnwidth]{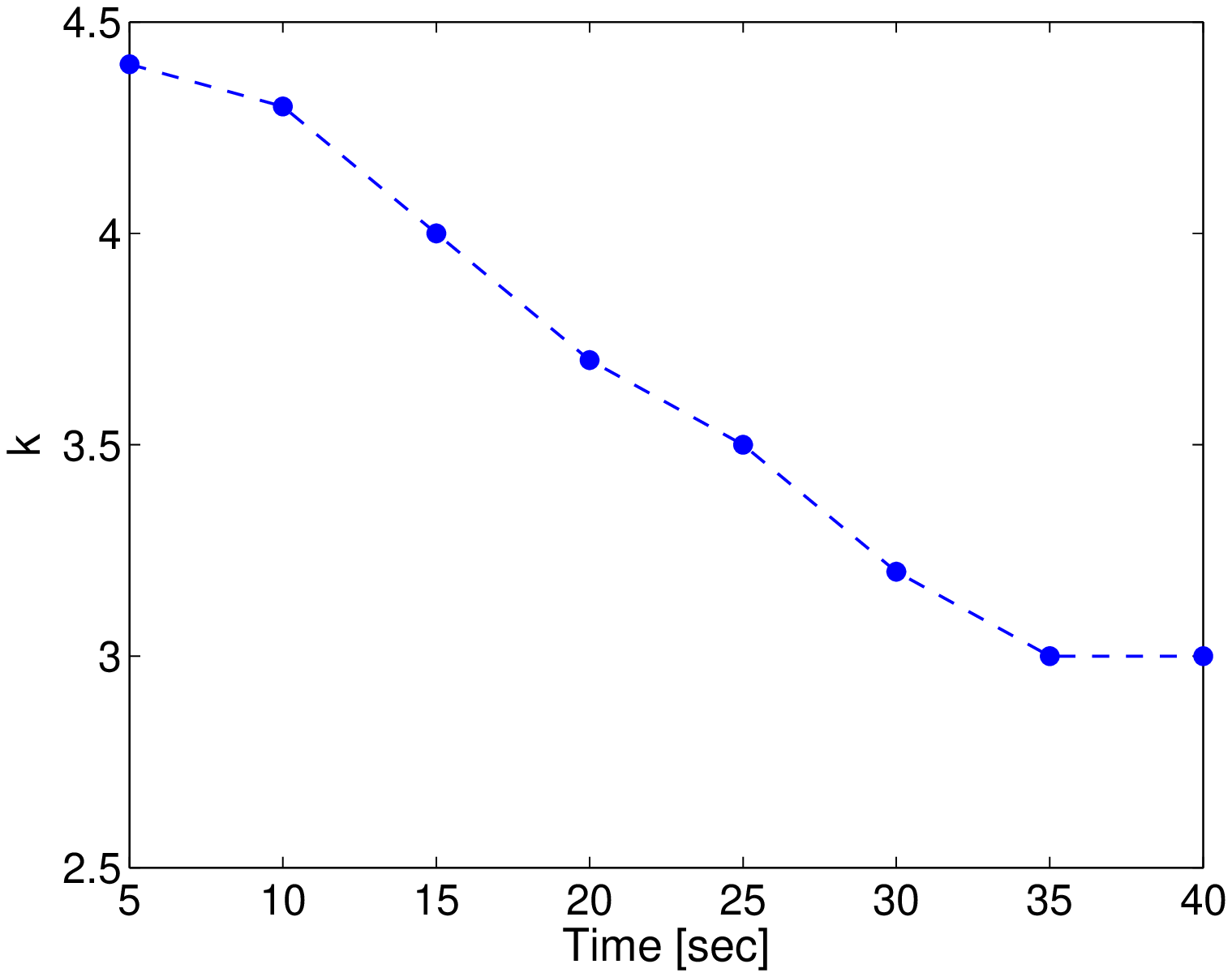}%
        \label{f:F2p:k}}\\
	\sidesubfloat[]{\includegraphics[width=0.40\columnwidth]{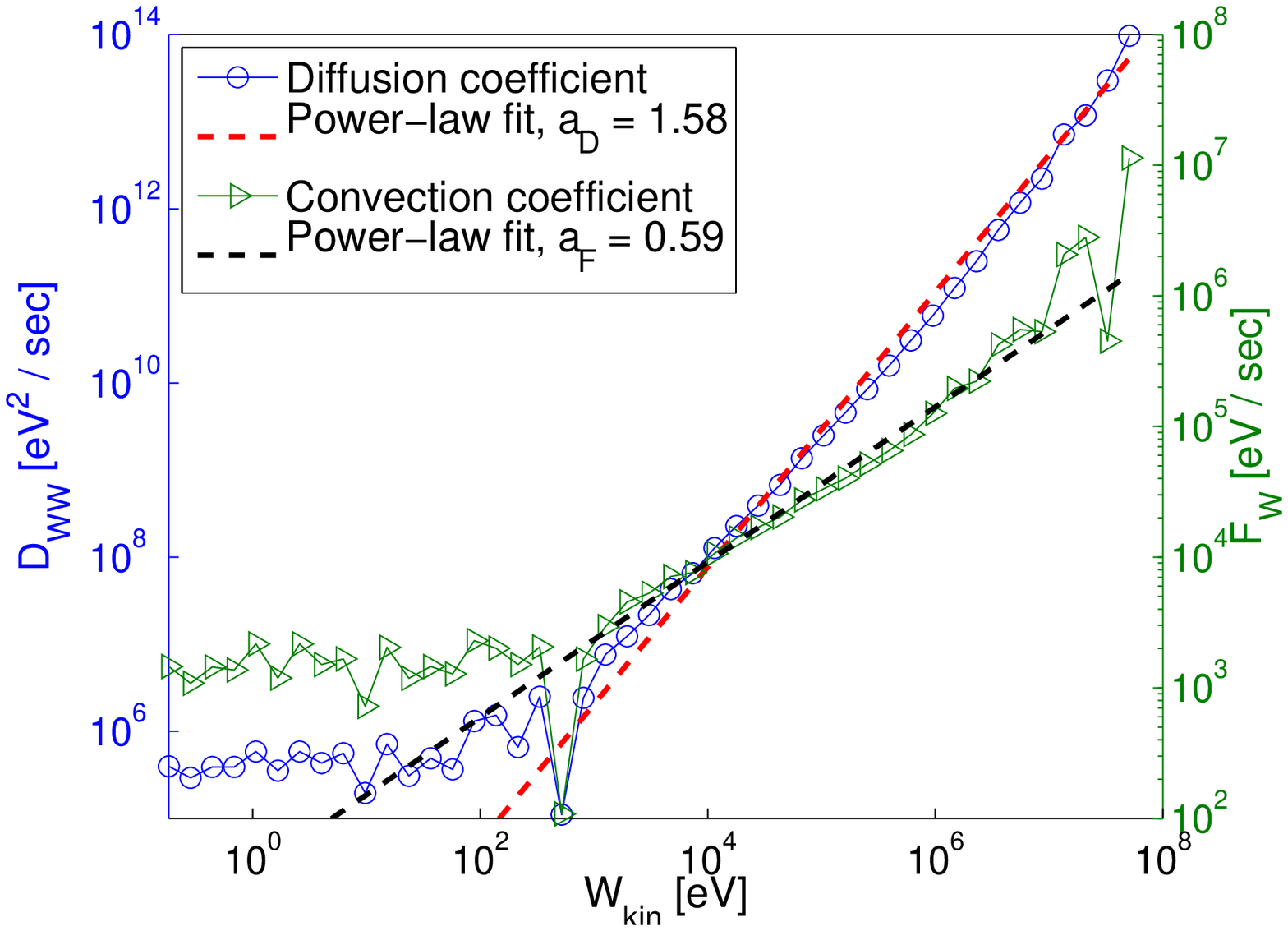}%
		\label{f:F2p:DF_W}}\hfill
	\sidesubfloat[]{\includegraphics[width=0.40\columnwidth]{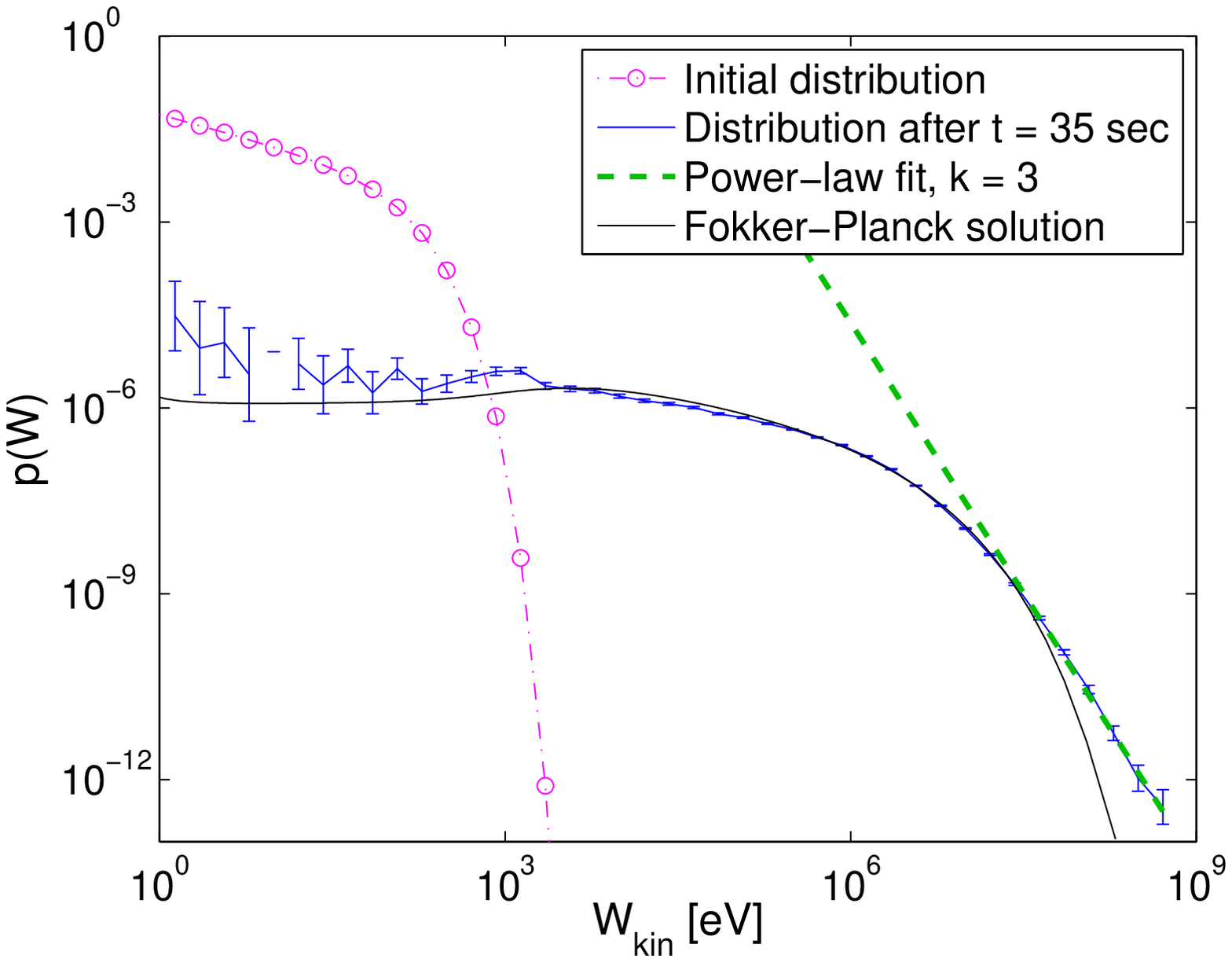}%
		\label{f:F2p:Wdistr}}
 	\caption{\protect\subref{f:F2p:mW} The kinetic energy of the electrons as a function of time (blue); their mean energy (black) with an exponential fit (green); the kinetic energy of three typical electrons. 
 	 \protect\subref{f:F2oI:mWlog} The temporal evolution of the mean kinetic energy of all the ions in the transient phase (the first seconds, black), together with a power-law fit ($\sim t^{2.1}$, green line).
    \protect\subref{f:F2p:k} The power-law index of the tail of the distribution at various times up to its stabilization. \protect\subref{f:F2p:DF_W} The energy diffusion and convection coefficients as functions of the kinetic energy. \protect\subref{f:F2p:Wdistr} Energy distribution at $t = 0$ and $t = \SI{35}{sec}$ for the electrons remaining inside the box, and the corresponding solution of the Fokker-Planck equation.}\label{f:F2p}
\end{figure}

Keeping the setup outlined above, we now impose periodic boundary conditions on the grid, i.e.~when an electron reaches any boundary of the grid, it re-enters from the corresponding grid-point at the opposite boundary. We only reduce $R$ down to $0.05$, since, due to the trapping in the acceleration volume, the energization is much more efficient and the electrons reach very quickly relativistic velocities. The electrons are monitored over a time-interval sufficiently large for the power-law index to stabilize, e.g.~for \SI{35}{sec} (see Fig.~\ref{f:F2p}).

The initial and the final distribution of the energy of the electrons after 35 seconds are shown in Fig.~\ref{f:F2p:Wdistr}. By this time the distribution can be considered ``asymptotic'', meaning that the power-law tail index has decreased to $k \approx 3$ and it stabilizes at that value, the system thus has the same characteristics as in the open-boundary case, see Fig.~\ref{f:F2o:Wdistr}.

The temporal evolution of the mean kinetic energy of the electrons and ions is similar to the open box case analysed above (see Fig.~\ref{f:F2p:mW}). The energy of the electrons increases exponentially after a short transient period (during which the mean energy follows a power law evolution over time, $<W> \sim t^{2.1}$ (see Fig.~\ref{f:F2p:mWlog}), as it is predicted for the non-relativistic particles by the hard sphere approximation), giving a numerical estimate of the acceleration time of $t_{\rm acc_{num}} \approx \SI{8}{sec}$, which is equal to $t_{\rm acc_{th}}$.

In Fig.~\ref{f:F2p:DF_W} the diffusion and convection coefficients at $t = \SI{35}{sec}$, as functions of the energy, are presented. They both exhibit a power-law shape for energies above \SI{1}{keV}, $D(W) = 37.98\ W^{1.58}$ and $F(W) = 38.57\ W^{0.59}$. The indices are thus very close to the ones of the power-laws in the open boundary case. We again use the estimated transport coefficients in the FP equation
{\bfseries (eq.~\ref{diff2}, for $t_{\rm esc}=\infty$), } 
solving it numerically as before, and the resulting energy distribution at final time is shown in Fig.~\ref{f:F2p:Wdistr}, which shows that we again have very good coincidence in the intermediate to high energy range, and some deviations in the highest energy range.

As in the open boundaries case discussed above, $\lambda_{\rm sc}$ is the key parameter which affects the evolution of the system. Keeping the characteristic length of the acceleration volume constant, and varying the density of the scatterers in the range $0.05 < R < 0.20$ (i.e.~$ \SI{3.3e8}{cm} < \lambda_{\rm sc} < \SI{8.3e7}{cm}$), the acceleration time decreases from $\approx \SI{8}{sec}$ to $\approx \SI{2}{sec}$ with increasing $R$,
and the power-law tail index $k$ decreases from $k\approx 3$ to $k\approx 2$.
%$3 \gtrsim k \gtrsim 2$.
The saturation time for the $k$-index lies within \num{25} and \SI{35}{sec}, so the quantitative behavior of the system depends indeed strongly on the mean free path of the interaction of the particles with the scatterers, $\lambda_{\rm sc}$.

%-----------------------------------------------------------------------------------
%-----------------------------------------------------------------------------------
\section{Discussion}\label{s:discussion}
Stochastic acceleration of electrons and ions by weak turbulence during solar flares has been discussed extensively in the astrophysical literature \cite[see the reviews by][]{Miller97, Petrosian12}. Several questions on the stochastic acceleration by weak turbulence still remain open. Let us mention a few here:
(a) The term ``turbulence'' in these studies means ``a spectrum of MHD waves with low amplitude ($\delta B/B\ll 1$)'' and the excited waves cover a specific range of wave numbers so that the resonant or transient acceleration of the particles by the waves can reach very high energies. The spectrum of the waves is assumed to be a power-law with specific slope. The excitation  or the formation by non linear wave-wave interaction of the specific spectrum, with the characteristics outlined above, through an explosive energy release in a complex magnetic topology remains an open question.
(b) The transport coefficients are estimated with the use of the quasilinear approximation, which is valid only for weak turbulence.
(c) The escape time has never been  quantitatively estimated, and it is used as a free parameter, calculated from the simple relation $t_{esc} \approx L/v,$ where $L$ is the characteristic size of the acceleration volume and $v$ the velocity of the energized particles.  Therefore, returning back to the three questions posed in the introduction, we conclude that: (1) The constraints posed for the validity of the theory of Stochastic Acceleration by Turbulence are very severe, and its relation with the original concept proposed by Fermi, the Stochastic Fermi Energization, is questionable. (2) We have shown that the escape time, which follows closely the acceleration time $t_{\rm acc}$ in the asymptotic state of the interaction of particles with AS, is an increasing function of the particle energy. This result is in agreement with recent observations \citep{Petrosian10}, although with some caution, due to  the relatively simple topology assumed by the authors in their analysis, which is contrary to the assumptions made in this article.	 (3) Based on the solution of the Fokker Planck equation and our numerical results, we have shown that when the transport coefficients are estimated properly the trapped particles sustain heating and acceleration of the plasma in a similar fashion with the non-trapped particles.

We have revisited in this article  the initial proposal put forward by \cite{Fermi49} and re-introduced the concept of ``magnetic clouds'' as ``Alfv\'enic Scatterers'', based on strong local magnetic fluctuations ($\delta B/B \approx 1$), formed at a small number of random places inside a complex magnetic topology and traveling with the Alfv\'en speed. Summing up the total energy carried by the random fluctuations ($U_{\delta B}$) and dividing by the energy carried by the ambient magnetic field $U_B$, the weak turbulent approximation ($U_{\delta B}/U_B\ll 1$) is still valid, but here now the energy is localized and carried by the moving coherent structures.
As we have shown in this article, the interaction of ions and electrons with the AS is a relatively simple and, at the same time, a very efficient mechanism for their heating and acceleration. Several aspects from our analysis can now be put in context in the energization of plasmas in the solar atmosphere during explosive events. The most important finding from our analysis is that the fast (on a time scale of about $t_{\rm acc}\approx$ \SIrange{10}{20}{sec}) heating and acceleration of the plasma particles depends on one parameter only, the mean free path of the particles interacting with the localized magnetic fluctuations $(\lambda_{\rm sc}).$ The energy distribution of the electrons reaches an asymptotic form with temperature around \SI{10}{keV} and a power-law tail with index $k \approx 2$ for $\lambda_{\rm sc} \approx \SI{e8}{cm}$.

Based on the above findings, we propose a very simple, four step scenario for the heating of the plasma and the acceleration of particles during the explosive phenomena on the Sun, which in our opinion solves a number of open questions that have been posed by current observations:

\begin{enumerate}
	\item  Step 1: ({\it Start of the magnetic instability and formation of Unstable Current Sheets})
	
	 The loss of equilibrium of large scale structures (based, as we outlined in the Introduction, on magnetic flux emergences from the convection zone, loss of stability of loops, the violent shuffling of magnetic field lines at the footpoints of closed loops) will initiate a major reconstruction of the large scale magnetic topology and  will cause the formation of Unstable Current Sheets of all scales \citep[see Fig.~\ref{f:cartoonb}, and][]{Gordovskyy2014}.
	
	\item Step 2: ({\it Large scale magnetic reconstruction and launch of a large scale MHD disturbance})
	
	The formation of complex magnetic topologies based on the violent reconstruction of the magnetic field  during the launch of a magnetic disturbance on all scales due to the formation of UCS inside the unstable magnetic configuration (see Fig. \ref{f:cartoonb}). Steps 1 and 2 are following the scenario proposed by \cite{Fletcher08} and are implemented numerically by \cite{Gordovskyy2014},  considering the deposition of energy in the low corona and upper chromosphere by large scale MHD disturbances during the collapse of a complex magnetic topology.

	\item Step 3: ({\it Formation of AS and UCS at random places inside the unstable magnetic topology})
	The formation of AS and UCS through the propagation of the large scale disturbances in the complex magnetic structures \citep[see Fig.~\ref{f:cartoona} and Fig.~5 in][]{Gordovskyy2014}.
		
	\begin{figure}[!ht]
        \sidesubfloat[]{\includegraphics[width=0.40\columnwidth]{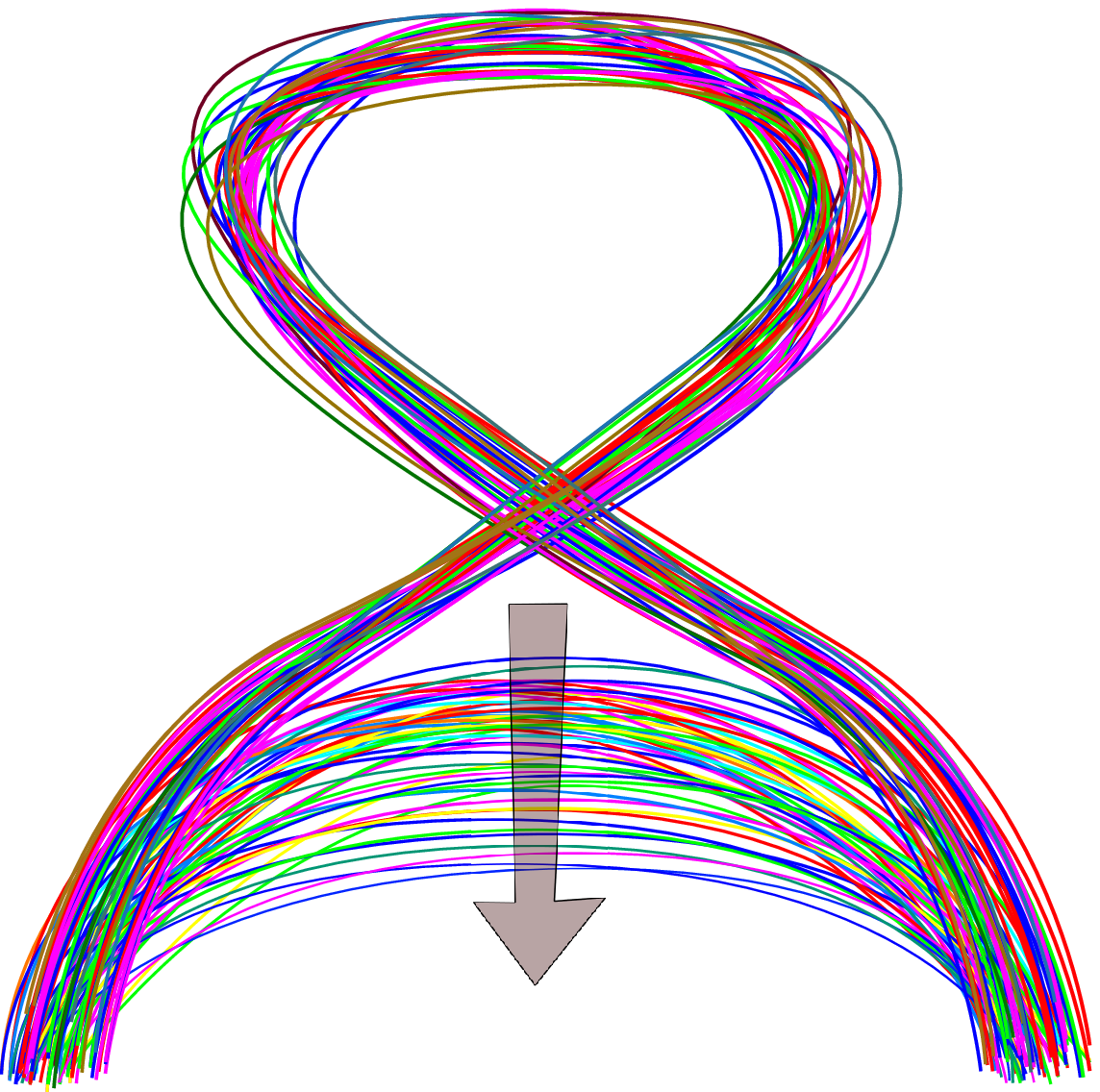}%
		    \label{f:cartoonb}}
	    \sidesubfloat[]{\raisebox{-5cm}{\includegraphics[width=0.50\columnwidth]{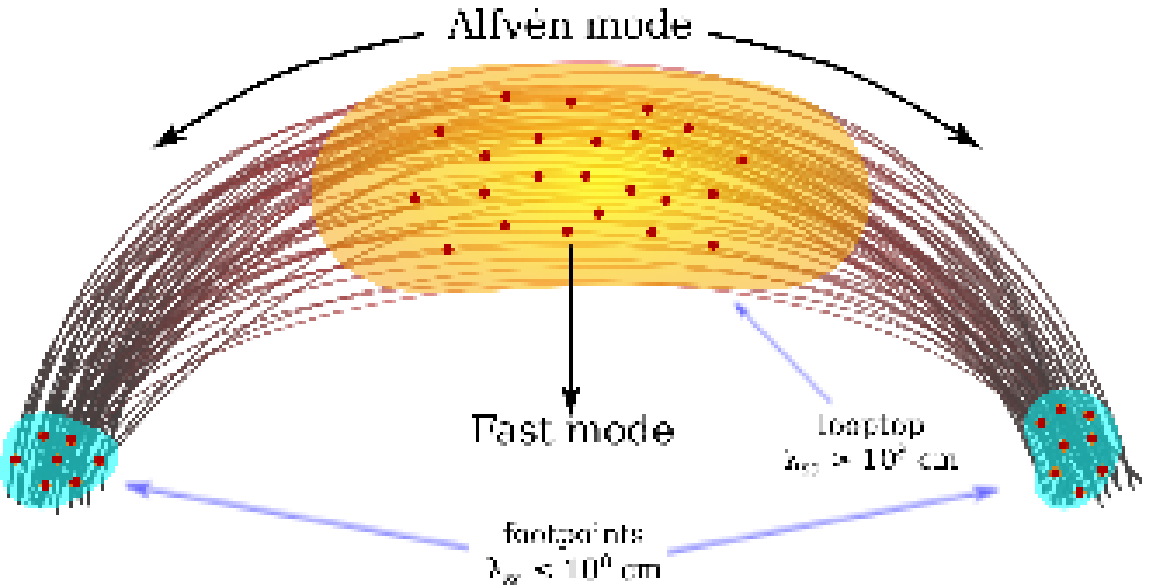}}%
            \label{f:cartoona}}
	    \caption{(a) The loss of stability of a large scale magnetic structure will initiate Unstable Current Sheets and launch a large scale disturbance, which will drive AS in the 3D topology associated with the unstable structure. (b) In the collapsing structures, AS (marked here with red dots) are formed inside the complex magnetic topology at the top and the footpoints, with very similar $\lambda_{\rm sc}$. The plasma at the top is intensely heated, particles are accelerated, and it is trapped for relatively long times. The footpoints are simultaneously energized through the AS formed by the MHD disturbance. }\label{f:cartoon}
    \end{figure}

	\item Step 4: ({\it loop top, foot points and chromospheric heating}) The AS and UCS formed in the closed magnetic topology (Fig.~\ref{f:cartoona}) 
	have  $\lambda_{\rm sc} > \SI{e8}{cm}$ at the loop top, where particles will be trapped, heated, and accelerated for relatively long times \citep[see][]{Krucker07, Krucker08}. More details for this process are presented in Sec.~\ref{periodic} about the AS.  The interaction of the plasma with the UCS has been discussed briefly by \cite{Vlahos16}. The MHD disturbance forming the AS will reach the two foot points simultaneously and since $\lambda_{\rm sc}$	is approximately the same at the footpoints and possibly smaller than at the loop top, the heating and the acceleration of particles will locally be more intense (see Sec.~\ref{Open} and Fig.~\ref{f:cartoonb})  and will compete favorably with the local Coulomb collisions, as it already does at the loop-tops (see Sec.~\ref{collisions} and Fig.~\ref{f:cartoona}). It would be interesting to analyse the propagation of the magnetic disturbance into the chromosphere and to study if the local heating can be efficient when $\lambda_{\rm sc}$ starts becoming comparable with the collisional mean free path (see Sec.~\ref{collisions}).
\end{enumerate}

The proposed scenario provides a way for the initial suggestions made  by Fermi to energize the solar plasma during explosive events without the need of a large number of particles to be transported from the coronal part of the unstable structure to the low corona and the upper chromosphere.

\section{Summary}\label{s:summary}
We developed a 3D lattice model where a small fraction of grid-points acts as ``scatterers'', in accordance with the initial suggestion of \cite{Fermi49} and the work of \cite{Parker58} and \cite{Ramaty79}. In our work, the emphasis is put on the coherent local fluctuations inside a complex magnetic topology, which are moving with the Alfv\'en speed, and which we here refer to as ``Alfv\'enic Scatterers'' (AS), and the initial Fermi approach is called Stochastic Fermi Energization (SFE).

The main results from our study are:
\begin{itemize}
	\item The Stochastic Fermi Energization (SFE) can reproduce the well known energy distribution of  astrophysical plasmas, where heating of the bulk and acceleration of the energetic particles co-exist.
	\item The density of the scatterers (or equivalently, the mean free path $\lambda_{\rm sc}$ of the interaction of the particles with the AS) controls the heating and the evolution of the energetic particles.
	\item The energy distribution reaches an asymptotic state on a time scale comparable to the acceleration time $t_{\rm acc}$. Similar results have been reported on the interaction of ions with a spectrum of Alfv\'en waves or electrons with a spectrum of whistler waves \citep[see][]{Miller90}. When the energy distribution reaches the asymptotic state, the mean escape time of the particles $t_{\rm esc}$ is comparable with $t_{\rm acc}$ {\bf if  $\lambda_{esc} \approx 10^8 cm.$}
	\item The index of the power-law of the particles in the energetic tail in the asymptotic state seems to agree very well with the simple formula derived by Fermi, $k=1+t_{\rm acc}/t_{\rm esc}$ 	and with the estimates of \cite{Parker58} and \cite{Ramaty79} for $W>>mc^2.$
	\item The escape time has a power-law dependence on the energy of the particles, $t_{\rm esc} \sim W^{0.3}$, which is different from the relation used in the analysis of the stochastic acceleration by waves, and it agrees  qualitatively  with observations \cite[see][]{Krucker07,  Petrosian10}. Yet, a more careful modeling is needed, both for the long trapping of the hard X-rays in the high corona \citep{Krucker07} and for the results reported by \cite{Petrosian10}.
	\item The transport coefficients are estimated from the dynamics of the particles interacting with the AS; the systematic acceleration coefficient has the form $F\sim W^{a_F}$ and the diffusion coefficient the form $D\sim W^{a_D}$, where $a_F = 0.59$ and $a_D = 1.51$,  quite close to the values predicted in the hard sphere approximation of \cite{Parker58} and \cite{Ramaty79}. The solution of the Fokker-Planck equation conforms well with the energy distribution derived from our numerical simulation of the dynamic evolution of the
	%energy distribution of
	particles, in the low and up to mildly-relativistic energy regimes.
	\item Collisions slow down the low energy electrons and ions and do not influence the power-law index in the asymptotic state.
	\item The trapped particles (periodic boundary conditions) are energized
	%the distribution
	much more efficiently than the particles in the open system. For this reason, we have used a longer mean free path, and the system reaches the asymptotic state on a longer time scale and the energy distribution of the energetic particles has a softer power-law index. The overall characteristics of the energy distribution are similar to the results reported for the open box. The transport coefficients were again estimated from the trajectories of the particles, and the solution of the Fokker-Planck equation agrees with the energy distribution estimated from the simulation of the dynamic evolution of the particles.
\end{itemize}

In summary, the formation of large scale {\bf local} magnetic fluctuation $(\delta B/B \approx 1)$ inside complex magnetic topologies during explosive phenomena forced us to reconsider the original SFE process. We note that the SFE process has only one free parameter, the mean free path $\lambda_{\rm sc}.$

Many question remain open and will be discussed in future publications. For example, what will happen to the system analyzed here if the SA are replaced by UCS \citep[see][]{Vlahos16}~? What will happen to a beam of particles interacting with the SA and being re-accelerated as they propagate inside the complex magnetic topology~? 
What will happen at a strong shock if the AS and UCS upstream and downstream of the shock surface are treated with the tools reported in this study~?
%-----------------------------------------------------------------------------------
%-----------------------------------------------------------------------------------
\begin{acknowledgements}
	We thank the anonymous referee for helpful comments and suggestions.
	 The authors acknowledge support by European Union (European Social Fund -- ESF) and Greek national funds through the Operational Program Education and Lifelong Learning of the National Strategic Reference Framework (NSRF) -- Research Funding Program: Thales: Investing in knowledge society through the European Social Fund.
\end{acknowledgements}
%-----------------------------------------------------------------------------------
%-----------------------------------------------------------------------------------

\end{document}